\documentclass[aps,prd,preprint,floats,epsf,nofootinbib]{revtex4-1}

\usepackage{graphicx}
\usepackage[caption=false]{subfig}
\usepackage{amsmath,amssymb,amsfonts,mathtools} 
\usepackage{slashed} 
\usepackage{braket} 
\usepackage{xcolor}
\usepackage{dsfont}
\usepackage{ulem}
\usepackage{dcolumn}
\usepackage{bm}
\usepackage{bbm}
\usepackage[hyperfootnotes=true]{hyperref}
\usepackage[T1]{fontenc}
\allowdisplaybreaks

\usepackage[utf8]{inputenc}

\usepackage{tikz}
\usetikzlibrary{calc}

\usepackage{tensor}
\usepackage{physics}

\graphicspath{{figures/}}

\begin{document}

\title{Deep Learning to Improve the Sensitivity of Di-Higgs Searches in the $4b$ Channel}

\author{\vspace{0.2cm}
{Cheng-Wei Chiang$^{\,1,2}$, Feng-Yang Hsieh$^{\,1}$, Shih-Chieh Hsu$^{\, 3}$, and Ian Low$^{\,4,5}$
}}
\affiliation{\vspace*{0.6cm}
$^1$\mbox{Department of Physics, National Taiwan University, Taipei 10617, Taiwan}\\
$^2$\mbox{Physics Division, National Center for Theoretical Sciences, Taipei 10617, Taiwan}\\
$^3$\mbox{Department of Physics, University of Washington, Seattle, WA 98195, USA}\\
$^4$\mbox{High Energy Physics Division, Argonne National Laboratory, Argonne, IL 60439, USA}\\
$^5$\mbox{Department of Physics and Astronomy, Northwestern University, Evanston, IL 60208, USA} 
\vspace*{0.5cm}}

\begin{abstract}
	The study of di-Higgs events, both resonant and non-resonant, plays a crucial role in understanding the fundamental interactions of the Higgs boson. In this work we consider di-Higgs events decaying into  four $b$-quarks and propose to improve the experimental sensitivity by utilizing a novel machine learning algorithm known as Symmetry Preserving Attention Network (\textsc{Spa-Net}) --  a neural network structure whose architecture is designed to incorporate the inherent symmetries in particle reconstruction tasks. We demonstrate that the \textsc{Spa-Net} can enhance the experimental reach over baseline methods such as the cut-based and the Deep Neural Networks (DNN)-based analyses. At the Large Hadron Collider, with a 14-TeV centre-of-mass energy and an integrated luminosity of 300 fb$^{-1}$, the \textsc{Spa-Net} allows us to establish 95\% C.L. upper limits in resonant production cross-sections that are 10\% to 45\% stronger than  baseline methods.  For non-resonant di-Higgs production, \textsc{Spa-Net} enables us to constrain the self-coupling that is 9\% more stringent than the baseline method.
\end{abstract}

\maketitle

\section{Introduction}
\label{sec:introduction}

\begin{figure}[t]
		\centering
		\subfloat[Triangle diagram]{
			\includegraphics[height=0.3\textwidth]{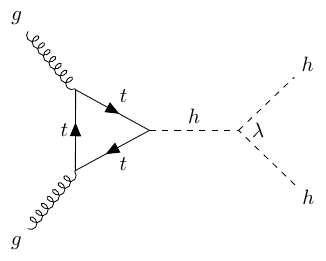}
		}
		\subfloat[Box diagram]{
			\includegraphics[height=0.3\textwidth]{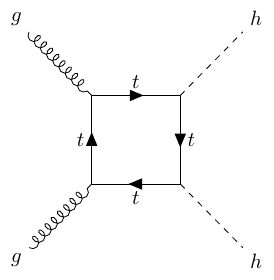}
		}
		\caption{The leading-order Feynman diagrams of the di-Higgs production in the SM.}
		\label{fig:feynman_diagrams_dihiggs_production_sm}
	\end{figure}

	Since the discovery of the 125-GeV Higgs boson, $h$, an immediate and pressing task has been to determine the Higgs potential and self-interactions, in addition to measuring its couplings to other SM particles, thereby verifying whether electroweak symmetry breaking (EWSB) is achieved in exactly the same way as the Standard Model (SM) prescribes~\cite{ATLAS:2022vkf,CMS:2022dwd}.  One important parameter in the Higgs potential is the trilinear coupling $\lambda$, which enters into the potential as:
	\begin{equation}
		V(h) = \frac{1}{2}m_h^2 h^2 + \lambda v h^3 + \lambda_{4h}h^4~.
	\end{equation}
	Here $m_h = 125$~GeV and the Higgs vacuum expectation value is $v=246$ GeV. In the SM, $\lambda=\lambda^{\rm SM}= m_h^2/(2v^2)$ and $\lambda_{4h} = \lambda_{4h}^{\rm SM}= m_h^2/(8v^2)$. Phenomenologically the Higgs trilinear coupling contributes to the pair production of Higgs bosons, which has not been observed experimentally. In the SM, the leading-order Feynman diagrams contributing to the di-Higgs production at a hadron collider, such as the Large Hadron Collider (LHC) at CERN, are shown in Fig.~\ref{fig:feynman_diagrams_dihiggs_production_sm}, where $\lambda$ features prominently.

	An experimental verification of the Higgs potential has important implications, as it is well known that the potential in the SM cannot induce a strong first-order electroweak phase transition in the early Universe, which is a crucial ingredient to explain the observed matter-antimatter asymmetry~\cite{Sakharov:1967dj, Kuzmin:1985mm, Cohen:1993nk}. It has been speculated that new physics should enter at an energy scale slightly higher than the weak scale to modify the Higgs potential at finite temperatures, to facilitate a sufficiently strong first-order phase transition.  In this case, the $\lambda$ coupling would be modified from its SM value, which could have a large impact on the di-Higgs production rate at the LHC~\cite{Kanemura:2004ch}. In addition, kinematic distributions of the Higgs pair could offer a unique window into new particles and new interactions above the weak scale~\cite{Dawson:2015oha, Chen:2014xra}.

	With a mass of 125~GeV, the Higgs boson predominantly decays into a $b\bar b$ pair. Therefore $hh\to 4b$ channel offers the largest rate among all possible decays of the Higgs pair which, nevertheless, suffers from the much larger background from multijets and multi-$b$'s production of QCD. Furthermore, because of our inability to distinguish a $b$-jet from a $\bar b$-jet, it is a very challenging task experimentally to form the correct pairing among the 4 $b$'s to reconstruct the Higgs mass and the associated kinematic distributions. In this work, we would like to propose employing machine learning algorithms from deep neural networks to help improve the sensitivity of experimental searches in the $4b$ channel. More specifically, we study the possibility of using a new neural network architecture called Symmetry Preserving Attention Network (\textsc{Spa-Net})~\cite{PhysRevD.105.112008, 10.21468/SciPostPhys.12.5.178, Fenton:2023ikr} to simultaneously perform signal/background separation and identify the correct pairings among the 4 $b$-jets in the final states. We will demonstrate that \textsc{Spa-Net} offers improved sensitivity over existing experimental techniques employed in the $4b$ channel \cite{ATLAS:2018rnh, ATLAS:2022hwc, ATLAS:2023qzf, CMS:2018qmt, CMS:2022cpr}, as well as over an analysis invoking the DNN machine learning algorithm \cite{Amacker:2020bmn}.

	We will consider two types of analyses in di-Higgs productions.  One involves the on-shell, resonant production of a  hypothetical new scalar particle, which subsequently decays into the Higgs pair.  As an explicit example, we consider the two-Higgs doublet models (2HDMs) in the alignment limit \cite{Carena:2013ooa, Carena:2014nza, Carena:2015moc}, where properties of the 125~GeV Higgs boson are SM-like. In this case, the Higgs pair is produced through the decay product of the heavy scalar \cite{Low:2020iua, Chen:2022vac}. We do not modify the SM contribution to di-Higgs production in this scenario. The other analysis, on the other hand, involves ``non-resonant'' production in the sense that we alter the SM trilinear coupling $\lambda$, of which the production cross-section is a function, and does not invoke direct production of new particles.

	This paper is organized as follows.  In Section~\ref{sec:sample_generation}, we describe the procedures employed for generating both signal and background samples utilized in the training of neural networks.  Section~\ref{sect:jetp} discusses three different jet pairing methods.  The first two are cut-based methods inspired by an ATLAS analysis, while the last is done with \textsc{Spa-Net}.  In Section~\ref{sec:selection_methods}, we provide the training procedures of DNN and \textsc{Spa-Net} classifiers and how to employ these neural network classifiers on the event selection task. In Section~\ref{sec:search_for_resonant_pair_production_of_higgs_bosons}, we perform the resonant analysis and discuss how \textsc{Spa-Net} can improve the sensitivity of the search. In Section~\ref{sec:higgs_self_coupling_constraints}, we build classifiers to discriminate non-resonant signal and background events and then demonstrate how \textsc{Spa-Net} performs better in constraining the Higgs trilinear coupling in comparison with more traditional methods. Finally, we summarize in Section~\ref{sec:conclusion}.


\section{Event generation}
\label{sec:sample_generation}

	  We use \verb|MadGraph5_aMC@NLO 3.3.1|~\cite{Alwall:2014hca} to generate both signal and background events at the centre-of-mass (CM) energy $\sqrt{s} = \text{13 TeV}$ with the \verb|NNPDF23_nlo_as_0119| PDF set~\cite{Ball:2012cx}. The LO matrix elements are considered. For parton showering and hadronization, we employ \verb|Pythia 8.306|~\cite{Sjostrand:2014zea} with \verb|NNPDF2.3 LO| PDF set. The detector simulation is performed using \verb|Delphes 3.5.0|~\cite{deFavereau:2013fsa}. Jets are reconstructed with \verb|FastJet 3.3.4|~\cite{Cacciari:2011ma} using the anti-$k_t$~\cite{Cacciari:2008gp} algorithm with radius $R = 0.4$. Only jets with a transverse momentum of $p_\text{T}\ge \text{20 GeV}$ are considered.

	\subsection{Signal Event Generation}
	\label{sub:signal_event_generation}

  	 	We consider two types of di-Higgs events:  resonant production through an on-shell new scalar boson and non-resonant production in the SM with, however, a rescaled Higgs self-coupling $\lambda = \kappa_\lambda \lambda^{\rm SM}$.

		For the resonant signal, Higgs boson pairs $hh$ are produced via the decay of heavy CP-even scalar $H$ in the 2HDM, which itself is produced through the gluon-fusion channel.  We consider  $m_H$ ranging from $\text{300 GeV}$ to $\text{1200 GeV}$. In this mass range, the $b$-jets can be reconstructed into four distinct energetic jets. The 2HDMC~\cite{Eriksson:2009ws} calculator with \verb|HiggsBounds-5.10.2|~\cite{Bechtle:2008jh, Bechtle:2011sb, Bechtle:2012lvg, Bechtle:2013wla, Bechtle:2015pma} and \verb|HiggsSignal-2.6.2|~\cite{Stal:2013hwa, Bechtle:2013xfa, Bechtle:2014ewa, Bechtle:2020uwn} extensions is used to compute the parameters at these benchmark points, which are submitted to \verb|MadGraph5_aMC@NLO 3.3.1| through the parameter card.

		The non-resonant signal is produced at one-loop at the leading order, via the Feynman diagrams shown in  Fig.~\ref{fig:feynman_diagrams_dihiggs_production_sm}. The process is simulated using the \verb|MadGraph5_aMC@NLO 3.3.1| with \verb+loop sm+ model. We leave $\kappa_\lambda$ as a free parameter varying over the domain of $[-10,15]$ when considering non-resonant productions.

		The $H\to hh, h \to b\overline{b}$ decays are implemented by \verb|MadSpin|~\cite{Artoisenet:2012st}. For resonant analysis, the $b$-tagging efficiency in \verb|Delphes| is modified based on the ATLAS MV2c10 $b$-tagger at the 70\% working point~\cite{ATL-PHYS-PUB-2016-012, ATLAS:2015thz}. At this working point, the light-jet (charm-jet) rejection is about 385 (12), which is the reciprocal of the false positive rate. For non-resonant analysis, the $b$-tagging efficiency is modified based on the ATLAS DL1r 77\% working point~\cite{ATLAS:2022qxm}. At this working point, the light-jet (charm-jet) rejection is about 130 (4.9).

		
	\subsection{Background Event Generation}
	\label{sub:background_event_generation}

		The main background is QCD multijet production: $pp \to b \overline{b} b \overline{b}$. The background for resonant and non-resonant analysis are simulated using different $b$-tagging settings as described in section \ref{sub:signal_event_generation}. The sub-leading background is top-quark pair production, which contributes to less than 10\% of the dominant background and is not included in the analysis. It's important to note that \verb|Pythia| considers the initial-state radiation (ISR) and final-state radiation (FSR) at the parton level. Therefore, there might be more than 4 jets after the jet clustering.
        
        In generating both the signal and background events, we implement a basic ``four-tag cut'', which requires at least four $b$-tagged $R = 0.4$ anti-$k_t$ jets with $p_\text{T} > \text{40 GeV}$ and pseudorapidity $\eta$ within the range $\abs{\eta} < 2.5$. The numbers of events passing this cut are given in Table~\ref{tab:samplesizes}.
        
		\begin{table}[t]
			\centering
			\caption{Sizes of various samples used for neural network study in resonant and non-resonant analyses. Each category consists of an equal size of signal and background samples.}
			\label{tab:samplesizes}
			\begin{tabular}{lccc}
				\hline \hline
							 & ~Training~ & ~Validation~ & ~Testing~ \\ \hline
				Resonant     & 950k     & 50k        & 100k    \\
				Non-resonant & 171k     & 9k         & 18k     \\
				\hline \hline
			\end{tabular}
		\end{table}

\section{Jet Pairing}
\label{sect:jetp}

    To reconstruct two Higgs boson candidates, we use three different jet assignment methods. The first two, $\Delta R + \text{min-}D_{hh}$ and $\text{min-}\Delta R$, are cut-based and inspired by the ATLAS analysis in Ref.~\cite{ATLAS:2018rnh, ATLAS:2023qzf}. The third one makes use of the \textsc{Spa-Net} neural network~\cite{PhysRevD.105.112008, 10.21468/SciPostPhys.12.5.178, Fenton:2023ikr}, a novel architecture specifically designed for the jet assignment task.

	\subsection{Cut-based Pairing}
	\label{sub:cut_based_pairing}

		In the $\Delta R + \text{min-}D_{hh}$ pairing method, the four $b$-jets with the highest $p_\text{T}$ are paired to construct two Higgs boson candidates. There are three possible pairings for the jets. Only the pairing that satisfies the following $\Delta R$ requirements is accepted:
		\begin{equation}\label{eq:DeltaR}
			\begin{aligned}
				\left.
				\begin{array}{c}
					\dfrac{\text{360 GeV}}{m_\text{4j}} - 0.5 < \Delta R_{\text{jj}}^{\text{lead}} < \dfrac{\text{653 GeV}}{m_{\text{4j}}} + 0.475 \\
					\dfrac{\text{235 GeV}}{m_\text{4j}}  < \Delta R_{\text{jj}}^{\text{subl}} < \dfrac{\text{875 GeV}}{m_{\text{4j}}} + 0.35 
				\end{array} 
				\right\} \text{ if } m_{\text{4j}} <  \text{1250 GeV} \\
				\left.
				\begin{array}{c}
					0 < \Delta R_{\text{jj}}^{\text{lead}} < 1 \\
					0 < \Delta R_{\text{jj}}^{\text{subl}} < 1 
				\end{array} 
				\right\} \text{ if } m_{\text{4j}} >  \text{1250 GeV}
			\end{aligned}
		\end{equation}
		where the $\Delta R_{\text{jj}}^{\text{lead}}$ is the angular distance between the jets in the $p_{\text{T}}$-leading Higgs boson candidate and $\Delta R_{\text{jj}}^{\text{subl}}$ for the sub-leading candidate, and $m_{\text{4j}}$ is the total invariant mass of the four jets. The angular distance is calculated using the formula $\Delta R = \sqrt{\Delta\eta^2 + \Delta\phi^2}$, where $\Delta\eta$ and $\Delta\phi$ represent the pseudorapidity and azimuthal angle differences between the two jets, respectively.

		If no pairing satisfies the above $\Delta R$ requirements, the event is dropped. If more than one pairing satisfies the $\Delta R$ requirements, we choose the one with the minimum $D_{hh}$, defined as 
		\begin{equation}
			D_{hh} = \frac{\abs{m_{h_1} - \frac{120}{110} m_{h_2}}}{\sqrt{1 + \left( \frac{120}{110} \right)^2}}\ ,
		\end{equation}
		where $m_{h_1}, m_{h_2}$ are the masses of the leading Higgs candidate and sub-leading Higgs candidate, respectively. The quantity $D_{hh}$ is the distance from $(m_{h_1}, m_{h_2})$ to the line connecting $(\text{0 GeV}, \text{0 GeV})$ and $(\text{120 GeV}, \text{110 GeV})$. The values of 120 GeV and 110 GeV account for energy loss.

		In the $\text{min-}\Delta R$ pairing method, the four $b$-tagged jets with the highest $p_{\text{T}}$ are used to form the two Higgs boson candidates. The $\text{min-}\Delta R$ method selects the pairing configuration in which the higher-$p_{\text{T}}$ jet pair has the smallest $\Delta R$ separation.


	\subsection{\textsc{Spa-Net} Pairing}
	\label{sub:spanet_pairing}

    	In this subsection, we provide an overview of the \textsc{Spa-Net}'s model structure and explain how this architecture is particularly well-suited for the jet assignment/pairing task. We also describe our approach to constructing the training samples for \textsc{Spa-Net}. Our goal is to train \textsc{Spa-Net} to recognize which jets arise from the decay of a given Higgs boson, thereby identifying the correct pairing of the Higgs boson candidates.

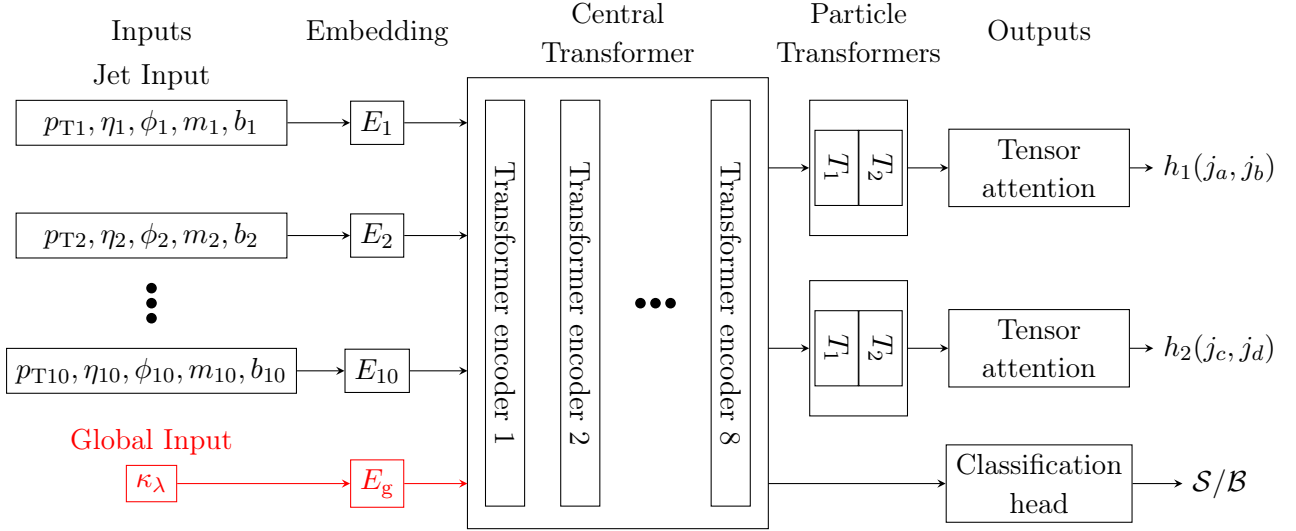
\begin{figure}[t]
    		\centering
    		{
\linespread{1.0}
\begin{tikzpicture}[>=stealth,scale = 1]

\pgfmathsetmacro{\h}{3.0}
\pgfmathsetmacro{\w}{2.0}

\coordinate (O) at (0,0);
\coordinate (Title) at (0, 1.2*\h);
\coordinate (Input) at (-3.1*\w,0);
\coordinate (Embedding) at (-1.6*\w,0);
\coordinate (CentralTransformer) at (0,0);
\coordinate (BranchEncoder) at (1.6*\w,0);
\coordinate (TensorAttention) at (2.8*\w,0);
\coordinate (Output) at (4.0*\w,0);
\coordinate (p1) at (-\w,-\h);
\coordinate (p2) at ( \w, \h);

\node[align=center] (CT) at (O |- Title) {Central\\Transformer};
\draw (-\w,-\h) rectangle (\w,\h);

\foreach \x/\n in {-0.75*\w/1, -0.25*\w/2, 0.75*\w/8}{
\node[draw, rotate=-90, minimum width=1.8*\h cm] at (\x,0) {Transformer encoder \n};
}
\foreach \i in {-0.2,0,0.2}{
	\fill ($(CentralTransformer)+(0.25*\w,0)$) +(\i,0) circle[radius=2pt];
}

\node (I) at (Input |- Title) {Inputs};
\node[draw,minimum width=1.8*\w cm] (I1) at ($(Input)+(0,0.8*\h)$) {$p_{\text{T}1}, \eta_1, \phi_1, m_1, b_1$};
\node[draw,minimum width=1.8*\w cm] (I2) at ($(Input)+(0,0.3*\h)$) {$p_{\text{T}2}, \eta_2, \phi_2, m_2, b_2$};
\node[draw,minimum width=1.8*\w cm] (I10) at ($(Input)+(0,-0.3*\h)$) {$p_{\text{T}10}, \eta_{10}, \phi_{10}, m_{10}, b_{10}$};
\node[draw,red] (Ig) at ($(Input)+(0,-0.8*\h)$) {$\kappa_\lambda$};

\node[above] at (I1.north) {Jet Input};
\node[above,red] at (Ig.north) {Global Input};

\foreach \i in {-0.2,0,0.2}{
	\fill (Input) +(0,\i) circle[radius=2pt];
}

\node[align=center] (E) at (Embedding |- Title) {Embedding};
\foreach \n in {1,2,10}{
	\node[draw] (E\n) at (I\n -| E) {$E_{\text{\n}}$};
	\draw[->] (I\n) -- (E\n);
	\draw[->] (E\n) -- (E\n -| p1);
}

\node[draw,red] (Eg) at (Ig -| E) {$E_{\text{g}}$};
\draw[->,red] (Ig) -- (Eg);
\draw[->,red] (Eg) -- (Eg -| p1);

\node[align=center] (BE) at (BranchEncoder |- Title) {Particle\\Transformers};
\node[draw,minimum width=0.65*\w cm,minimum height=0.6*\h cm] (B1) at ($(BranchEncoder)+(0, 0.6*\h)$) {};
\node[draw,minimum width=0.65*\w cm,minimum height=0.6*\h cm] (B2) at ($(BranchEncoder)+(0,-0.2*\h)$) {};
\node (BC) at ($(BranchEncoder)+(0,-0.8*\h)$) {};

\foreach \n in {1,2}{
	\node[draw, rotate=-90, minimum width=0.5*\w cm, above=2pt] at (B\n.west) {$T_{1}$};
	\node[draw, rotate=-90, minimum width=0.5*\w cm, below=2pt] at (B\n.east) {$T_{2}$};
}

\node[align=center] (TA) at (TensorAttention |- Title) {Outputs};
\node[draw,minimum width=1.2*\w cm, align=center] (TA1) at (TA|-B1) {Tensor\\attention};
\node[draw,minimum width=1.2*\w cm, align=center] (TA2) at (TA|-B2) {Tensor\\attention};
\node[draw,minimum width=1.2*\w cm, align=center] (CLS) at (TA|-BC) {Classification\\head};

\node (O1) at (Output|-TA1) {$h_1 (j_a, j_b)$};
\node (O2) at (Output|-TA2) {$h_2 (j_c, j_d)$};
\node (OC) at (Output|-CLS) {$\mathcal{S/B}$};

\foreach \i in {1,2}{
\draw[->] (p2|-B\i) -- (B\i);
\draw[->] (B\i) -- (TA\i);
\draw[->] (TA\i) -- (O\i);
}

\draw[->] (p2|-CLS) -- (CLS);
\draw[->] (CLS) -- (OC);
\end{tikzpicture}

}
    		\caption{The high-level model structure of \textsc{Spa-Net}. Each $E_i$ is an embedding layer, $T_i$ is the transformer encoder, and $h_i$ is the jet assignment result which contains two jets $j_i$ for the Higgs decay. The particle transformer is a stack of transformer encoders. The global input is only used in the non-resonant analysis.}
    		\label{fig:SPANet_structure}
    	\end{figure}

    	Figure~\ref{fig:SPANet_structure} shows the high-level model structure of \textsc{Spa-Net}~\cite{PhysRevD.105.112008, 10.21468/SciPostPhys.12.5.178, Fenton:2023ikr}. The embedding blocks encode the input features to the embedding vectors living in the latent space. These embedding vectors are fed into the central transformer, which is a stack of transformer encoders. The central transformer then outputs the event embedding vector, which is used in the jet assignment and the classification tasks. For the jet assignment, the event embedding vector is encoded by the particle transformers and the tensor attentions.  Finally, \textsc{Spa-Net} constructs the jet assignment results from these outputs. The network architecture has a feed-forward structure for the classification head.

    	There are several inherent symmetries in the jet assignment task. For example, the detector signatures of quarks and anti-quarks are nearly indistinguishable. Consequently, it is important to consider all possible combinations of jets originating from these partons. Moreover, the reconstruction task is insensitive to the swapping of specific labels. For instance, while a Higgs boson decays into a pair of $b$ and $\overline{b}$, inverting the labels would still result in the same reconstructed Higgs boson. Similarly, in the di-Higgs case, swapping the pairing results of two Higgs would lead to the same event reconstruction. The model structure of \textsc{Spa-Net} is designed to incorporate these symmetries.

    	Due to the properties of the transformer, the event embedding vector in \textsc{Spa-Net} is independent of the order of the input jets. Moreover, \textsc{Spa-Net} utilizes the technique of symmetric tensor attention~\cite{10.21468/SciPostPhys.12.5.178}, which constructs a tensor with permutation symmetries of labels ({\it e.g.}, the $b \bar b$ and $hh$ pairs). Therefore, the output also contains label permutation symmetries.  These properties enable \textsc{Spa-Net} to preserve the symmetry inherent to the jet assignment problems.

        In our context, it is essential to emphasize that \textsc{Spa-Net} is not restricted to using only the $b$-tagged jets for the jet assignment task, but considers all jets in an event.  This allows the network to make a correct prediction even in the situation where some of the jets are mistagged.  Therefore, \textsc{Spa-Net} can utilize a   larger dataset in the pairing task than the traditional methods.

    	The input features for the \textsc{Spa-Net} are a list of jets, each represented by their 4-vector $(p_\text{T}, \eta, \phi, m)$ as well as a boolean $b$-tag, where $\phi$ is the azimuthal angle and $m$ is the invariant mass. To be specific, we only keep the 10 highest $p_\text{T}$ jets in each event. For each event, we define the correct jet assignments by matching the jets to the simulated truth quarks within an angular distance of $\Delta R < 0.4$. In case a simulated truth quark is matched to more than one jet, such an event will be dropped. Furthermore, some simulated truth quarks may not be matched to any jet, in which case the event will not be used in training either. The percentage of samples satisfying these matching conditions to the samples passing the four-tag cut varies from 77\% to 89\%.  Note that these matching conditions are only required for training \textsc{Spa-Net}. However, in the final analysis, \textsc{Spa-Net} is applied to all events passing the four-tag cut.

\section{Neural Network Classifiers}
\label{sec:selection_methods}

    After jet pairing, the next step is to distinguish the signal from the background. In addition to a cut-based approach, two types of neural network classifiers are employed in this study: Dense Neural Network (DNN), which is a conventional deep learning architecture used as the baseline neural network approach, and \textsc{Spa-Net}, which could also be used as a classifier to separate signal from background.

	To perform a DNN-based analysis, we construct a DNN classifier to distinguish between signal and background events. We implement our DNN using the \verb|Tensorflow|~\cite{tensorflow2015-whitepaper} library. The network consists of simple dense layers and the internal node uses the rectified linear unit (ReLU) as the activation function. The categorical cross-entropy is used as the loss function, which is then minimized by the \verb|Adam| algorithm. Hyperparameters of the DNN are selected by utilizing the \verb|Optuna|~\cite{akiba2019optuna} hyperparameter optimization package. The learning rate, hidden dimension, and the number of layers are optimized by performing 100 iterations of hyperparameter optimization, and the set of hyperparameters that produces the best classification accuracy is selected for full training.

    Upon training, the DNN is used to determine whether an event is a signal or a background. The DNN classifier assigns a signal score $p_{\text{signal}}$ to every event, which represents the confidence that this event is a signal. An event is classified as a signal if its $p_{\text{signal}}$ is larger than $p_{\text{th}}$, a threshold score determined through the maximization of sensitivity $S / \sqrt{B}$, where $S$ and $B$ represent the number of signal and background events, respectively.

    For the \textsc{Spa-Net} classifier, it is important to note that the \textsc{Spa-Net} classification head does not take the results from the jet assignment part. Using the transformer outputs alone produces better performance compared to including the jet assignment results because errors in the jet assignment part can affect the overall performance. However, even if we only use the results from the classification head, we still train both the jet assignment and classification tasks simultaneously. The reason is that the jet assignment part can help \textsc{Spa-Net} build the embedding space structure. As a result, training on both tasks allows us to achieve better performance while using the same size of training samples compared to only training on the classification head.

    The hyperparameters of \textsc{Spa-Net} are selected using the \verb|Optuna| hyperparameter optimization package. We optimize the learning rate, dropout rate, gradient clipping, L2 penalty, hidden dimension, number of encoder layers, number of branch encoder layers, and number of classification layers. Each set of hyperparameters is trained for 10 epochs. We perform 100 iterations of hyperparameter optimization, and the set of hyperparameters that produces the best classification accuracy is selected for full training.

    Following the hyperparameter optimization process, \textsc{Spa-Net} is trained for 50 epochs using the \verb|AdamW| optimizer with L2 regularization applied to all parameter weights. The total loss in \textsc{Spa-Net} combines the contributions from both jet assignment and classification parts, indicating that these parts are not trained independently. The loss from each part is computed separately and summed with equal weights.

    Similarly, the \textsc{Spa-Net} assigns a signal score $p_{\text{signal}}$ to each event. To select the di-Higgs candidate events, we set a requirement $p_{\text{signal}} > p_{\text{th}}$ and the threshold $p_{\text{th}}$ is determined through the maximization of sensitivity $S / \sqrt{B}$.


\section{Search for resonant Di-Higgs Production}
\label{sec:search_for_resonant_pair_production_of_higgs_bosons}

	For resonant Higgs boson pairs, we describe the steps to set the 95\% confidence level (CL) upper limits on the cross-section of the resonant production of a new heavy scalar $H$ decaying into two Higgs bosons $hh$ and demonstrate that \textsc{Spa-Net} gives the best limit among all three methods.

	\subsection{Event selection in resonant analysis}

    	\subsubsection{Cut-based selection}
    	\label{subs:cut_based_selection_in_resonant_analysis}

            After jet pairing as described in Section \ref{sect:jetp}, we define the leading Higgs boson candidate $h_1$ to be the one with the highest scalar sum of jet $p_\text{T}$. The sub-leading Higgs is denoted by $h_2$. The following transverse momentum cuts are further applied to the leading and sub-leading Higgs candidates \cite{ATLAS:2018rnh}:
    		\begin{equation}
    			\begin{aligned}
    				p_{\text{T}}^{\text{lead}} &> m_{\text{4j}} \times 0.5 - \text{103 GeV}~, \\
    				p_{\text{T}}^{\text{subl}} &> m_{\text{4j}} \times 0.33 - \text{73 GeV}~, \\
    			\end{aligned}
    		\end{equation}
    		where  $m_{\text{4j}}$ is the total invariant mass of the two Higgs candidates,  $p_{\text{T}}^{\text{lead}}$ is the transverse momentum of the leading Higgs boson candidate, and $p_{\text{T}}^{\text{subl}}$ is for the sub-leading Higgs boson candidate.

    		For background rejection, we first apply a cut on the pseudorapidity difference between the two Higgs candidates $\abs{\Delta\eta_{hh}} < 1.5$.	 Next, we define the quantity $X_{hh}$~\cite{ATLAS:2018rnh}
    		\begin{equation}\label{eq:signal_region}
    			X_{hh} = \sqrt{\left( \frac{m_{h_1} - \text{120 GeV}}{0.1 m_{h_1}} \right)^2 + \left(\frac{m_{h_2} - \text{110 GeV}}{0.1 m_{h_2}} \right)^2}\ ,
    		\end{equation}
    		where $m_{h_1}$ is the mass of the leading Higgs candidate, and $m_{h_2}$ is the mass of the sub-leading Higgs candidate. Events with $X_{hh} < 1.6$ are considered as in the signal region. The reference masses of 120~GeV and 110~GeV account for energy losses in the detector.

    		To suppress the $t\overline{t}$ background, a top veto cut is needed.  We form ``$W$ candidates'' by pairing every possible pair of jets with $p_\text{T} > \text{40 GeV}$ and $\abs{\eta} < 2.5$, including those that are not selected as the $h$ candidates.  We then build ``top quark candidates'' by pairing the $W$ candidates with each remaining jet selected for the $h$ candidates. For each possible top quark candidate, we calculate the quantity $X_{Wt}$ defined as \cite{ATLAS:2018rnh}
    		\begin{equation}
    			X_{Wt} = \sqrt{\left( \frac{m_{W} - \text{80 GeV}}{0.1 m_{W}} \right)^2 + \left(\frac{m_{t} - \text{173 GeV}}{0.1 m_{t}} \right)^2} 
    		\end{equation}
    		where $m_W$ is the mass of the $W$ candidate, and $m_t$ is the mass of the top quark candidate. Events with the smallest $X_{Wt} < 1.5$ are vetoed.


    	\subsubsection{DNN selection}
    	\label{subs:dnn_selection_in_resonant_analysis}

        	To use the DNN classifier, we need to first apply jet assignments. We employ the $\text{min-}\Delta R$ and \textsc{Spa-Net} pairing methods to construct the Higgs candidates and generate two separate training datasets. Subsequently, we utilize these datasets to train two separate DNN classifiers and use the signal scores to separate the signal from the background. The input variables utilized by the DNN classifiers are summarized in Table~\ref{tab:DNN_variables_resonant}, as inspired by Ref.~\cite{Amacker:2020bmn}. These features include the 4-vector of the two Higgs candidates, the angular distance $\Delta R$ between the two jets associated with each Higgs candidate, the $b$-tagging information of the four jets, and the transverse momentum of the di-Higgs system.

    		\begin{table}[t]
    			\centering
    			\caption{Input variables for the dense neural network.}
    			\label{tab:DNN_variables_resonant}
    			\begin{tabular}{lcc}
    			\hline\hline
    				Reconstructed objects       & Input variables   & ~\#~ \\ \hline
    				Higgs candidate             & $(p_\text{T}, \eta, \phi, m)$ & 8  \\
    				Jet                         & $\Delta R(j_1,j_2)$                    & 2  \\
    				$b$-tagging                 & Boolean for $j_i \in h_{1,2}^{\text{cand}}$       & 4  \\
    				Di-Higgs system             & $p_\text{T}^{hh}, m_{hh}$        & 2 \\
    			\hline\hline
    			\end{tabular}
    		\end{table}


    	\subsubsection{\textsc{Spa-Net} selection}
    	\label{subs:spanet_selection_in_resonant_analysis}

    		Similar to the DNN selection approach, in the \textsc{Spa-Net} approach we utilize the signal scores generated by \textsc{Spa-Net} and set a specific threshold value. Events with a signal score greater than or equal to this threshold will be considered candidate resonant events. The threshold values are determined through the maximization of the sensitivity $S / \sqrt{B}$.

	
	\subsection{Results from resonant searches}
	\label{sub:resonant_results}

\begin{figure}[t]
			\centering
			\includegraphics[width=0.7\textwidth]{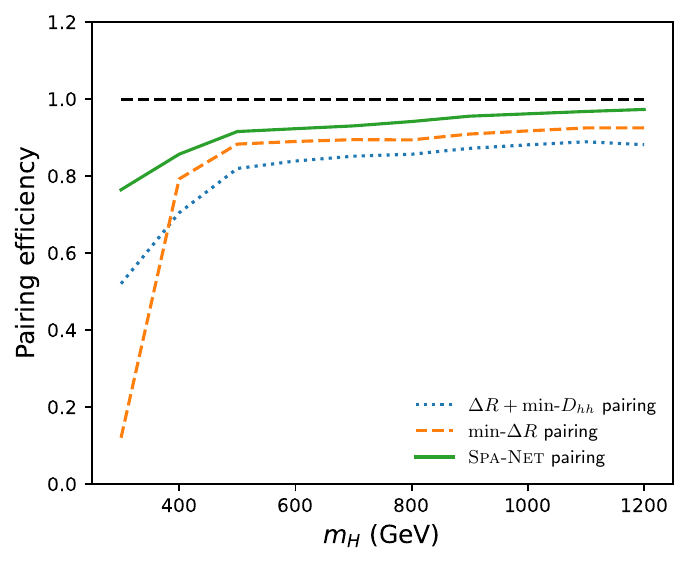}
			\caption{The pairing performance for samples with different $m_H$.}
			\label{fig:pairing_performance_mH}
		\end{figure}

		For the cut-based and DNN selection methods, it is necessary to construct the Higgs boson candidates. Figure~\ref{fig:pairing_performance_mH} shows the pairing efficiency of various methods. All pairing methods exhibit better performance in the higher mass region, while the pairing efficiency declines more significantly as the mass goes below $\sim 500$~GeV. This effect is especially noticeable in the $\text{min-}\Delta R$ method. This is because, in the low-resonance region, the Higgs boson obtains lower energy, causing the $b$-jet pair to have a larger $\Delta R$ separation, resulting in reduced performance. The \textsc{Spa-Net} pairing method outperforms other methods for all mass values.

\begin{figure}[th]
			\centering
			\includegraphics[width=0.7\textwidth]{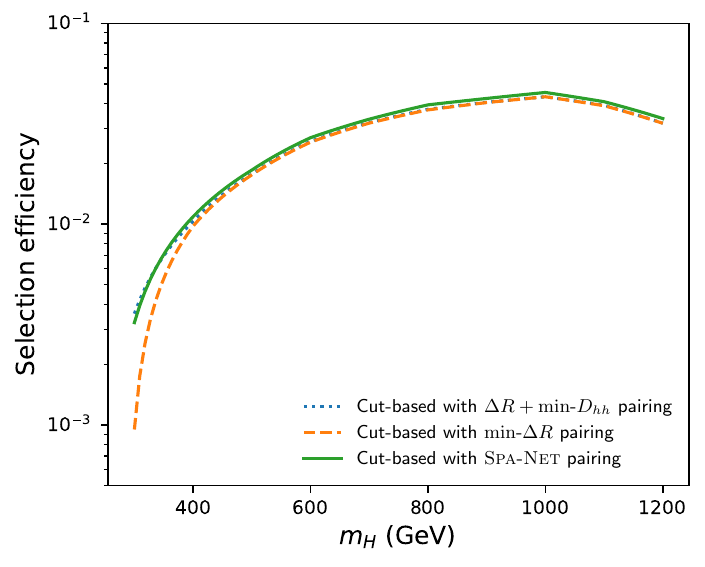}
			\caption{The selection efficiency for samples with different $m_H$. 
			}
			\label{fig:selection_efficiency}
		\end{figure}

		Figure~\ref{fig:selection_efficiency} shows the selection efficiency, defined to be the ratio of the number of events that pass the final cut to the total number of events without applying any cuts, for the cut-based selection with different pairing methods. The corresponding selection efficiencies of background samples range from $9.4 \times 10^{-5}$ to $2.82 \times 10^{-4}$. All three pairing methods exhibit similar performance. In the low-mass region, the efficiency is reduced due to the lower energy. The $\text{min-}\Delta R$ pairing method has even lower efficiency in this region due to its inferior pairing performance.

		Table~\ref{tab:resonant_classification_results_summary} presents the training results for the neural network classifiers, where we use the accuracy (ACC) and the area under the Receiver Operating Characteristic (ROC) curve (AUC) as two evaluation metrics. The DNN classifier with the \textsc{Spa-Net} pairing method shows better performance than DNN with the $\text{min-}\Delta R$ pairing method. The \textsc{Spa-Net} classifier has the best performance among the three classifiers. The difference between the DNN and the \textsc{Spa-Net} classifiers arises from the input features. While the DNN employs well-known physical observables as the input features, the \textsc{Spa-Net} classifier uses event embedding vectors, which cannot be readily interpreted as traditional physical observables. Nonetheless, these event embedding vectors seem better suited for the event classification task. To understand the physical information encoded in these event embedding vectors, further analysis is needed to find out the relationship between the high-level physical observables and the components of the event embedding vectors.

		\begin{table}[htpb]
			\centering
			\caption{The classification performance of different neural network classifiers. The ACC and AUC are evaluated based on 10 trainings.}
			\label{tab:resonant_classification_results_summary}
			\begin{tabular}{lcc}
			\hline\hline
			Classifier          & ~~~ACC~~~   & ~~~AUC~~~   \\ \hline
			DNN with $\text{min-}\Delta R$ pairing & $0.865 \pm 0.001$ & $0.938 \pm 0.001$ \\
			DNN with \textsc{Spa-Net} pairing      & $0.876 \pm 0.001$ & $0.946 \pm 0.001$ \\
			\textsc{Spa-Net}          			   & $0.894 \pm 0.002$ & $0.961 \pm 0.001$ \\
			\hline\hline
			\end{tabular}			
		\end{table}

\begin{figure}[t]
			\centering
			\includegraphics[width=0.7\textwidth]{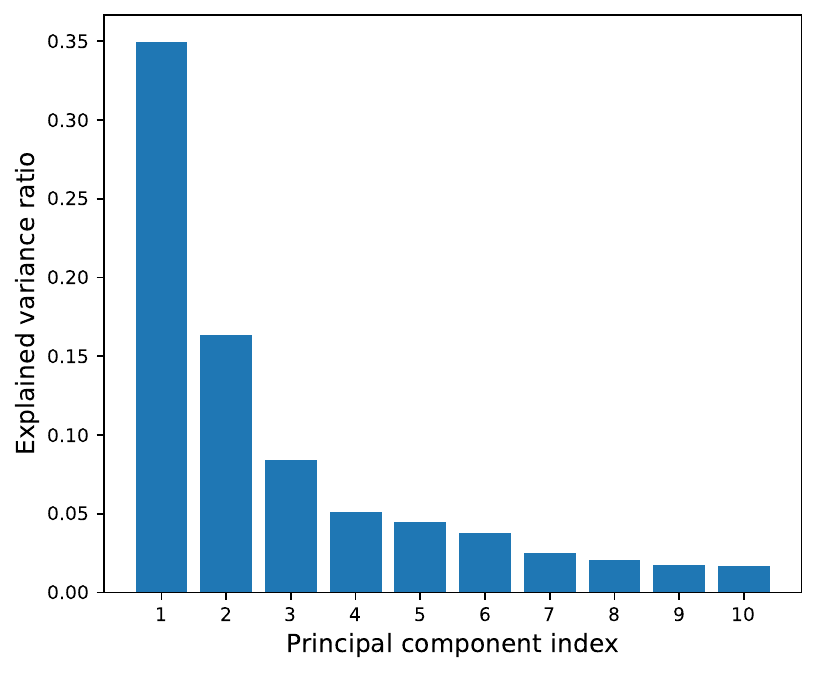}
			\caption{The variance ratio of the first ten principal components in resonant analysis.}
			\label{fig:PCA_variance_ratio_resonant}
		\end{figure}

		To better understand the embedding vectors, we have performed the Principal Component Analysis (PCA), which uses an orthogonal linear transformation that transforms the data to a new basis. (The PCA class from the \verb|scikit-learn|~\cite{scikit-learn} package is used.) In the new basis, the components are ordered by their variance. Figure~\ref{fig:PCA_variance_ratio_resonant} shows the variance importance for the first 10 principal components. The first three components can explain about 60\% of the total variance. This indicates that these components capture significant information from the event embedding vectors. Therefore, in the following analysis, we employ  only these first three principal components.

\begin{figure}[t]
			\centering
			\includegraphics[width=0.95\textwidth]{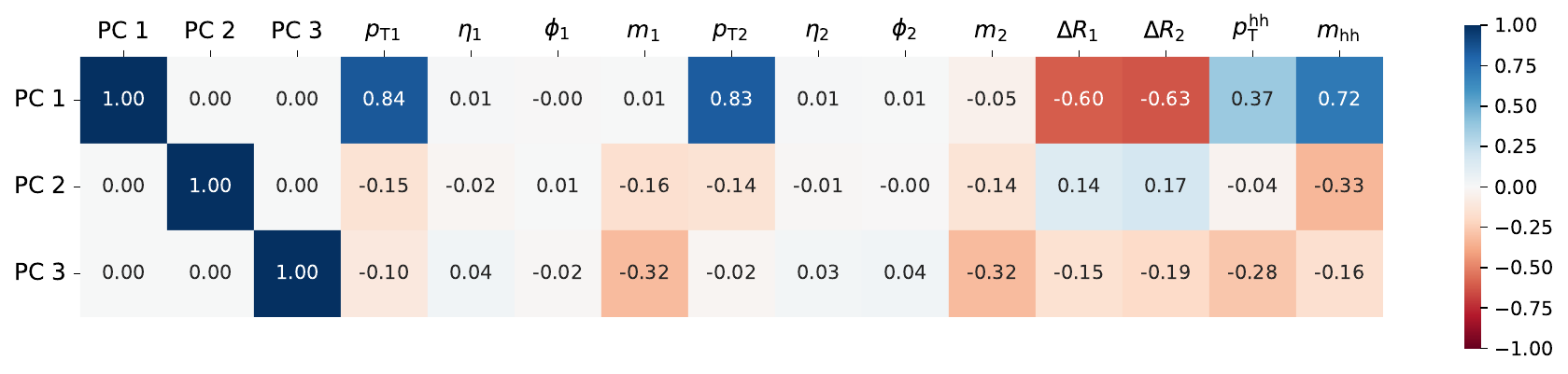}
			\caption{The correlation coefficients between the first three principal components and high-level observables.}
			\label{fig:correlation_coefficients_pca-resonant}
		\end{figure}

		To find out the correlation between the first three principal components and selected DNN input features, we have calculated the correlation coefficients, with the results shown in Figure~\ref{fig:correlation_coefficients_pca-resonant}. The features such as Higgs candidates' $p_{\text{T}}$, $\Delta R$, di-Higgs system's $p_{\text{T}}^{hh}$, and $m_{hh}$ have higher correlation coefficients with the principal components. This suggests that the event embedding vectors indeed contain important and meaningful physical information. Also, the transformer block in \textsc{Spa-Net} can learn and extract relevant physical information from the input features and encode it into event embedding vectors.

		Even though there is a high correlation between the embedding vector and high-level observables, the better training results of \textsc{Spa-Net} imply that the embedding vector contains additional information. This information might not correspond to the familiar high-level physical parameters, which nonetheless proves effective for the classification task. These findings highlight the fact that the \textsc{Spa-Net}  can extract the relevant physical information from the input data efficiently. In addition, the flexibility of \textsc{Spa-Net} allows it to construct more suitable variables for the classification task. As a consequence, the training performance exceeds that of utilizing only well-known physical parameters.

		To obtain the upper limits on the cross-section of resonant Higgs boson pair production, we use the reconstructed $m_{hh}$ distribution. From the binned $m_{hh}$ distribution, we can obtain the likelihood function $L$, which consists of a product of Poisson distributions for the number of events in each bin.

		The signal strength $\mu$ is chosen as the parameter of interest (POI). The profile likelihood ratio takes the following form:
		\begin{equation}
			-2 \Delta \ln L(\mu) \equiv -2 \ln \left( \frac{L (\mu)}{L(\hat{\mu})} \right)			
		\end{equation}
		where $\hat{\mu}$ is the maximum likelihood estimate of $\mu$. The upper limit on the signal strength is computed by the package \verb|pyhf|~\cite{pyhf,pyhf_joss}, which is based on the $\text{CL}_{\text{s}}$ method~\cite{Read:2002hq}. The POI is excluded at the 95\% CL when $\text{CL}_{\text{s}}$ is less than $0.05$. Then we can convert the results to the upper limit of the cross section.

\begin{figure}[t]
			\centering
			\includegraphics[width=0.7\textwidth]{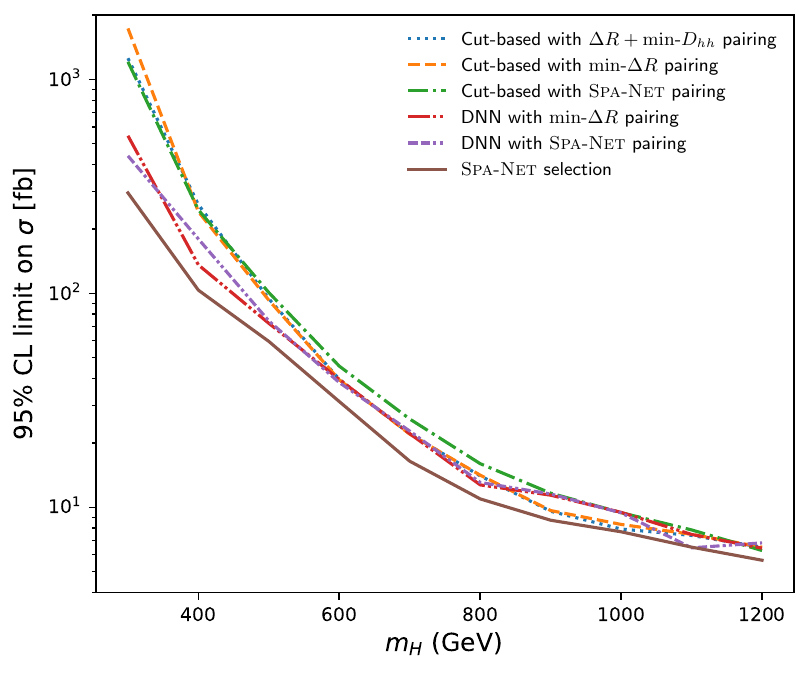}
			\caption{The upper limit on the cross-section of resonant di-Higg production for different $m_H$.}
			\label{fig:CL_limit_resonant}
		\end{figure}

		In setting the upper limit, we consider a luminosity of $\mathcal{L} = \text{300 fb}^{-1}$ for the 14-TeV LHC. Since the kinematics of the 13-TeV and 14-TeV samples are similar, we can scale the cross sections to those of the 14-TeV samples. Figure~\ref{fig:CL_limit_resonant} shows the upper limits on the resonant Higgs pair production as a function of $m_H$ for different selection methods. While all methods give similar results in the high resonance region, the \textsc{Spa-Net} selection has the most superior performance, providing the most stringent constraints throughout the considered mass range. The DNN selection methods give higher upper limits than \textsc{Spa-Net}, and the cut-based selection methods give the worst results. Specifically, \textsc{Spa-Net} selection enables us to establish cross-section upper limits that are 10\% to 45\% stronger compared to DNN with $\text{min-}\Delta R$ pairing.

\begin{figure}[th]
			\centering
			\includegraphics[width=0.9\textwidth]{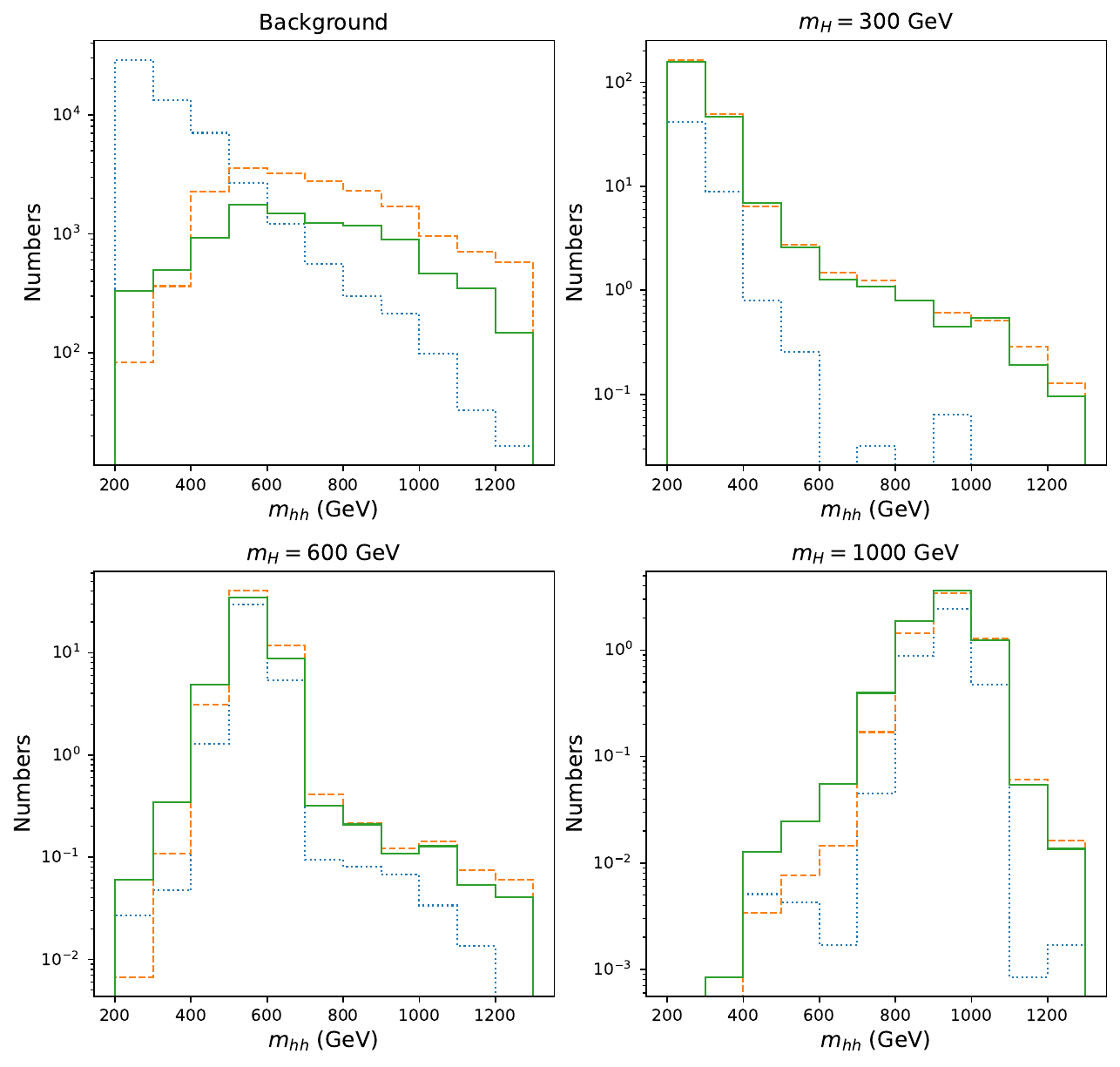}
			\caption{The invariant mass $m_{hh}$ distribution for the cut-based selection with \textsc{Spa-Net} pairing (blue dotted bins), the DNN with \textsc{Spa-Net} pairing (orange dashed bins), and the \textsc{Spa-Net} selection (green solid bins). The number of events is normalized to the target luminosity $\mathcal{L} = \text{300 fb}^{-1}$. Other pairing methods would give similar results to the corresponding selection method with \textsc{Spa-Net} pairing.}
			\label{fig:mhh_distribution}
		\end{figure}

		Figure~\ref{fig:mhh_distribution} shows the invariant mass $m_{hh}$ distributions using various selection methods. In the high-mass region, the DNN and \textsc{Spa-Net} selection methods let more background and signal events pass. Consequently, the results are similar to those obtained by the cut-based selection. In the low-mass region, the \textsc{Spa-Net} selection method can cut more background events, which accounts for why the \textsc{Spa-Net} selection achieves more stringent upper limits in this specific mass range.

\section{Higgs self-coupling constraints}
\label{sec:higgs_self_coupling_constraints}

	In this section, our signal events consist of non-resonant Higgs boson pairs with different Higgs self-coupling scale factor $\kappa_\lambda$. Similar to the analysis on resonant production, we compare the cut-based method with two distinct neural network classifiers, the DNN and the \textsc{Spa-Net},  which are used to identify candidate signal events. We will show that the \textsc{Spa-Net} architecture is better suited for the event classification task, and the improved classification results can yield stronger constraints on the coupling scale factor $\kappa_\lambda$.

    \subsection{Event selection in non-resonant analysis}

    	\subsubsection{Cut-based selection}
    	\label{subs:cut_based_selection_in_non_resonant_analysis}

    		The cut-based event selection methods are similar to that of the resonant analysis. In the non-resonant analysis, we utilize the $\text{min-}\Delta R$ and \textsc{Spa-Net} pairing methods. The high-level physical variables of two Higgs boson candidates are used in subsequent selection steps and analysis.

    		To further reduce the multijet background, we apply a cut on the pseudorapidity difference between the two Higgs candidates $\abs{\Delta\eta_{hh}} < 1.5$. To reduce the $t\overline{t}$ background, we employ the top veto cut. We compute the quantity $X_{Wt}$ and the events with the smallest $X_{Wt} < 1.5$ are vetoed.
    		
    		Finally, events with $X_{hh} < 1.6$ are considered as in the signal region (see Eq.~(\ref{eq:signal_region}) for the definition of $X_{hh}$). Those events would be used to determine the Higgs self-coupling constraints. 
    				

    	\subsubsection{DNN selection}
    	\label{subs:dnn_selection_in_non_resonant_analysis}

    		The signal samples are the non-resonant samples with different $\kappa_\lambda$ values. The input features of the network are summarized in Table \ref{tab:DNN_variables}, as inspired by Ref.~\cite{Amacker:2020bmn}. To obtain these input features, the Higgs pairing needs to be determined first. We use the $\text{min-}\Delta R$ pairing method and the \textsc{Spa-Net} pairing method to construct the Higgs candidates and generate two separate training datasets, which are then used to train two different DNN classifiers. For training, we set $\kappa_\lambda = [-5, -3, -1, 1, 2, 3, 5, 7, 9, 12]$ and generate samples for each value in the list. The signal events used in the training consist of different $\kappa_\lambda$ samples. For each $\kappa_\lambda$, the same number of sample events is used. For background samples, since they lack the specific $\kappa_\lambda$ information for the input feature, a $\kappa_\lambda$ value is randomly chosen from the above list.  We use the same sizes of signal and background samples as the scheme given in Table~\ref{tab:samplesizes} in the neural network training.

    	    \begin{table}[t]
    			\centering
    			\caption{Input variables for the dense neural network.}
    			\label{tab:DNN_variables}
    			\begin{tabular}{lcc}
    			\hline\hline
    				Reconstructed objects       & ~Variables used for training~   & ~\#~ \\ \hline
    				Higgs candidate             & $(p_\text{T}, \eta, \phi, m)$ & 8  \\
    				Jet                         & $\Delta R(j_1,j_2)$                    & 2  \\
    				Missing transverse momentum & $E_{\text{T}}^{\text{miss}}, \phi(\vec{p}_{\text{T}}^{\text{miss}})$  & 2  \\
    				Leptons                     & $N_e, N_\mu $                  & 2  \\
    				$b$-tagging                   & Boolean for $j_i \in h_{1,2}^{\text{cand}}$       & 4  \\
    				Di-Higgs system             & $p_\text{T}^{hh}, m_{hh}$        & 2  \\
    				Self coupling               & $\kappa_\lambda$        & 1  \\
    			\hline\hline
    			\end{tabular}		
    		\end{table}
    

    	\subsubsection{\textsc{Spa-Net} selection}
    	\label{subs:spanet_selection_in_non_resonant_analysis}

            Here we utilize \textsc{Spa-Net} to perform both the jet assignment and signal/background classification tasks. The input data contains information about the reconstructed jets and the global event features. \textsc{Spa-Net} would output both the jet pairing and the type of the events (signal v.s. background). In contrast to the DNN classifier, the key advantage is that \textsc{Spa-Net} can process the jet-level information and can preserve the permutation symmetry inherent in the problem.

    		The input features contain a list of jets, where each jet is represented by its 4-vector $(p_\text{T},\eta,\phi, m)$ as well as a Boolean $b$-tag. Additionally, we input the self-coupling scale factor $\kappa_\lambda$ as the global feature of each event. The sample preparation follows the same procedure as described for the DNN case in the previous section.
    		

	\subsection{Non-resonant analysis results}
	\label{sub:non_resonant_analysis_results}

		\begin{figure}[t]
			\centering
			\includegraphics[width=0.7\textwidth]{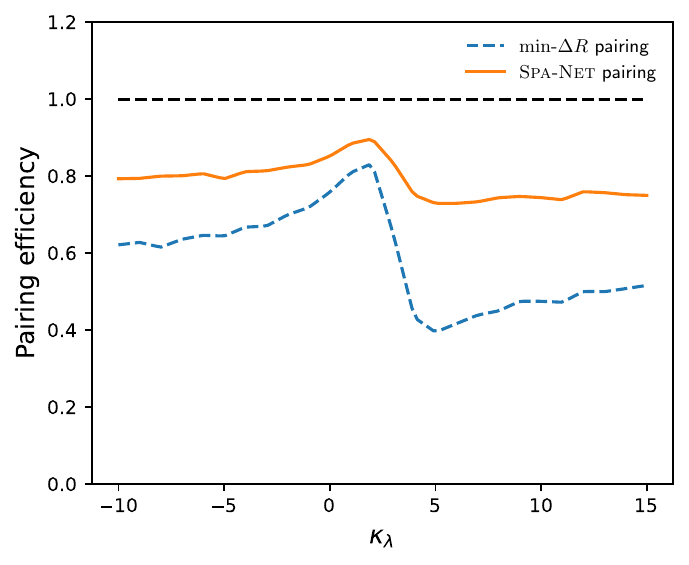}
			\caption{The pairing performance for different $\kappa_\lambda$ samples. The \textsc{Spa-Net} is trained on pairing and classification tasks at the same time.}
			\label{fig:pairing_performance_kappa}
		\end{figure}

\begin{figure}[t]
			\centering
			\includegraphics[width=0.7\textwidth]{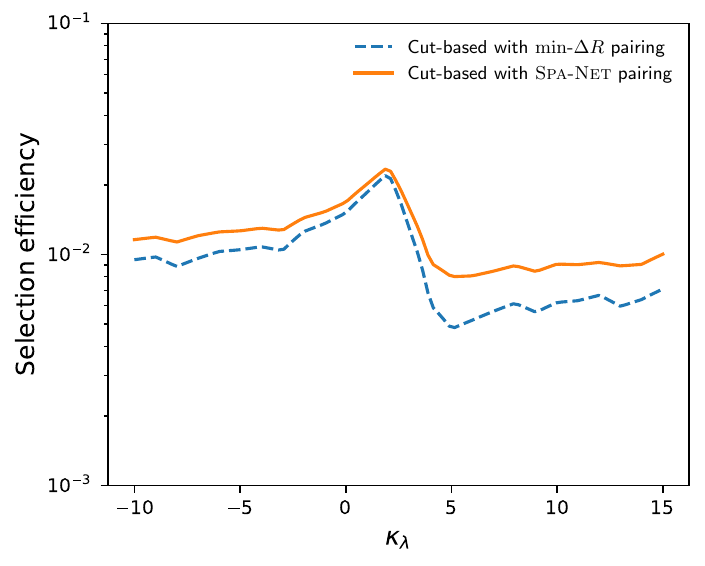}
			\caption{The selection efficiency for different $\kappa_\lambda$ samples. The selection efficiency is the ratio of the number of events that pass the final cut to the total number of events without applying any cuts. The corresponding selection efficiencies of background samples range from $9.7 \times 10^{-5}$ to $2.72 \times 10^{-4}$.}
			\label{fig:non_resonant_selection_efficiency}
		\end{figure}

        To obtain the high-level physical observables for cut-based selection and DNN training, we need to construct the Higgs boson candidates with different jet assignment algorithms. Figure~\ref{fig:pairing_performance_kappa} shows the pairing efficiency of the two jet pairing methods:  $\text{min-}\Delta R$ and \textsc{Spa-Net}. Both methods exhibit their best performance around $\kappa_\lambda = 2$. For the $\text{min-}\Delta R$ method, the pairing efficiency ranges from 40\% to 80\% while for \textsc{Spa-Net} the pairing efficiency ranges from 70\% to 90\%. Therefore, the \textsc{Spa-Net} pairing method outperforms the $\text{min-}\Delta R$ method for all coupling values. Figure \ref{fig:non_resonant_selection_efficiency} shows the selection efficiency for the cut-based selection using different pairing methods. The curves of the selection efficiency are similar to the ones of pairing efficiency, meaning that the selection efficiency and the pairing results are highly correlated.

		\begin{table}[t]
			\centering
			\caption{The classification performance of different selection methods.  The ACC and AUC are evaluated based on 10 trainings.}
			\label{tab:classification_results_summary}
			\begin{tabular}{lcc}
			\hline\hline
			Classifier          & ~~~~~~ACC~~~~~~   & ~~~~~~AUC~~~~~~   \\ \hline
			DNN with $\text{min-}\Delta R$ pairing~ & $0.799 \pm 0.011$ & $0.881 \pm 0.012$ \\
			DNN with \textsc{Spa-Net} pairing	    & $0.803 \pm 0.004$ & $0.884 \pm 0.004$ \\
			\textsc{Spa-Net}          			    & $0.828 \pm 0.002$ & $0.911 \pm 0.001$ \\
			\hline\hline
			\end{tabular}			
		\end{table}

		In Table~\ref{tab:classification_results_summary}, we present the training results for the classifiers. The DNN classifiers show similar performance for both pairing methods, while the \textsc{Spa-Net} classifier has the best performance among all. To better understand the event embedding vectors and find out the relationship between the input features of DNN and those of \textsc{Spa-Net}, we perform a similar analysis as in the resonant case.

\begin{figure}[t]
			\centering
			\includegraphics[width=0.7\textwidth]{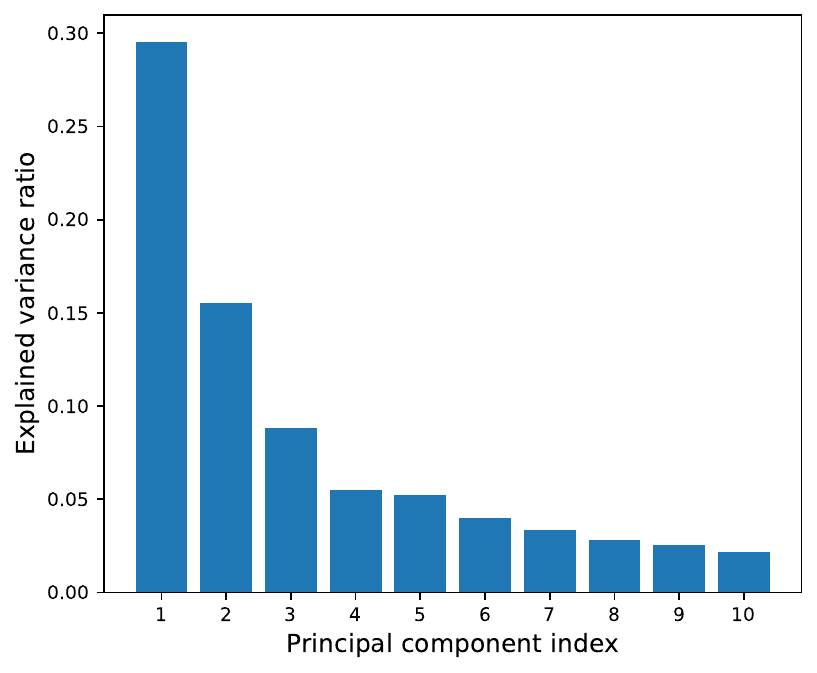}
			\caption{The variance ratio of the first ten principal components in the non-resonant analysis.}
			\label{fig:PCA_variance_ratio-nonresonant}
		\end{figure}

		First, we performed the PCA on the event embedding vectors. Figure~\ref{fig:PCA_variance_ratio-nonresonant} shows the variance importance for the first 10 principal components, with the first three components being able to explain about 50\% of the total variance. Since these components capture significant information from the event embedding vectors, we focus on these first three principal components in subsequent analysis.

\begin{figure}[t]
			\centering
			\includegraphics[width=0.95\textwidth]{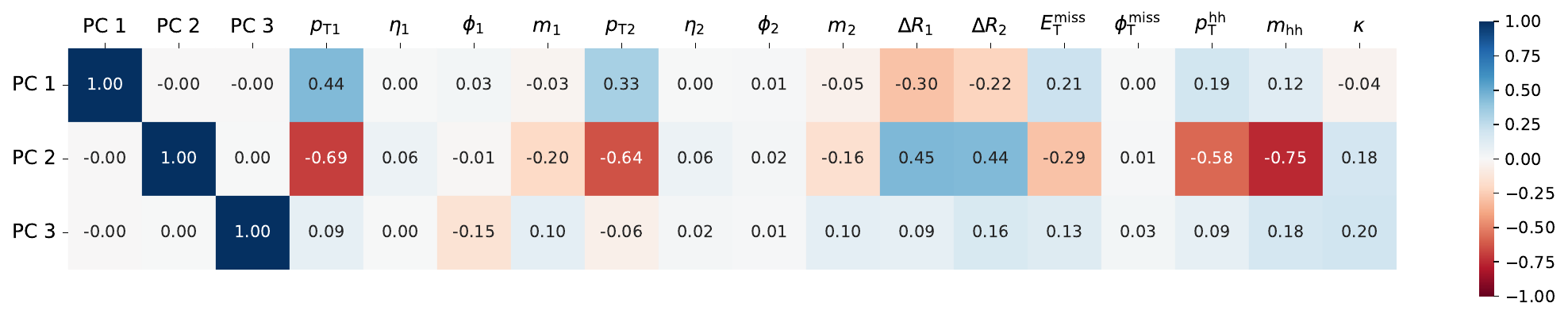}
			\caption{The correlation coefficients between the first three principal components and high-level observables.}
			\label{fig:correlation_coefficients_pca-signal-background}
		\end{figure}

		Next, we compute the correlation coefficients between these principal components and high-level physical observables. The results are shown in Figure \ref{fig:correlation_coefficients_pca-signal-background}. The features such as Higgs candidates' $p_{\text{T}}$, $\Delta R$, di-Higgs system's $p_{\text{T}}^{hh}$, and $m_{hh}$ have higher correlation coefficients with the principal components. These results are similar to the resonant case. These findings suggest that the event embedding vectors indeed contain meaningful physical information. Moreover, the flexibility of \textsc{Spa-Net} enables it to explore beyond the familiar physical parameters, thereby allowing \textsc{Spa-Net} to obtain superior performance compared to the traditional DNN structure in classification tasks.

		For the $\kappa_\lambda$ constraints setting, we consider a luminosity of $\mathcal{L} = \text{300 fb}^{-1}$ and use the 14-TeV cross-sections. Since the kinematics of the 13-TeV and 14-TeV samples are similar, we can scale the cross sections to those of the 14-TeV samples.

		Similar to the resonant analysis, to obtain constraints on the Higgs self-coupling scale factor $\kappa_\lambda$, we use the reconstructed $m_{hh}$ distribution, from which we compute the likelihood function $L$ consisting of a product of Poisson distributions for the number of events in each bin. The signal strength $\mu$ is chosen as the POI. The values of the coupling $\kappa_\lambda$ are excluded at the 95\% CL if the predicted cross-section of the signal model with that configuration is excluded with $\text{CL}_{\text{s}}<0.05$. Alternatively, we can obtain the exclusion limit by using the profile likelihood with the coupling $\kappa_\lambda$ as POI. A scan of the profile likelihood ratio is taken as a function of the coupling, and from this we can set the $1.96\sigma$-level constraints.

		\begin{figure}[t]
			\centering
            \includegraphics[width=0.7\textwidth]{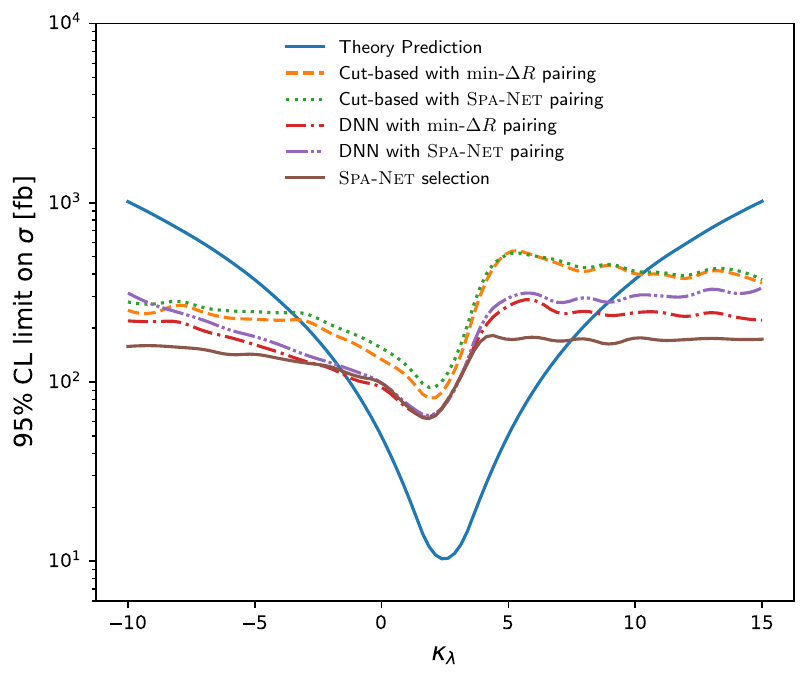}
			\caption{The upper limits of the cross-section with different $\kappa_\lambda$. The theory prediction is the cross section computed from \texttt{MadGraph5\_aMC@NLO}. The coupling $\kappa_\lambda$ with the cross-section greater than the upper limit would be excluded.}
			\label{fig:CL_upper_limit}
		\end{figure}

		\begin{figure}[th]
			\centering
            \includegraphics[width=0.7\textwidth]{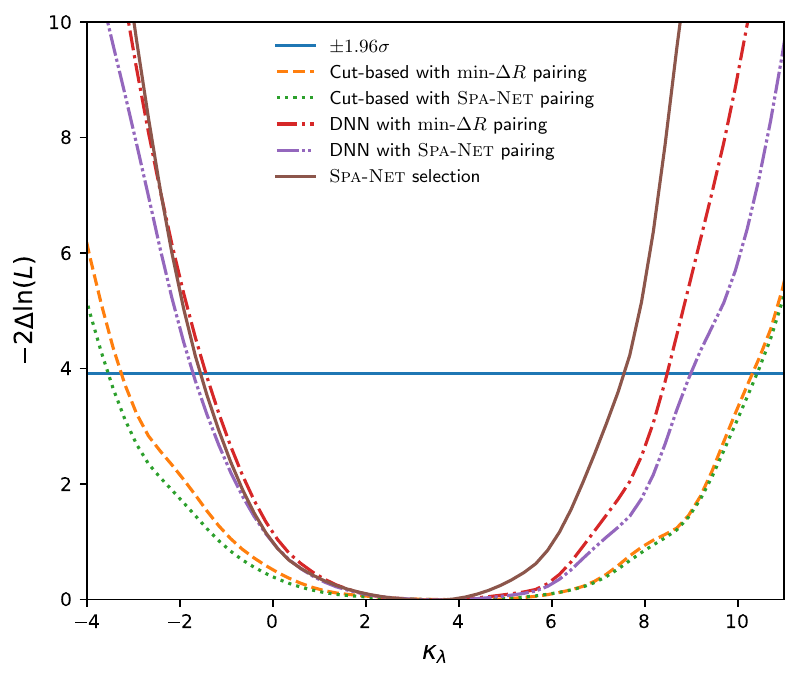}
			\caption{The profile likelihood ratio scans for $\kappa_\lambda$. The blue line indicates the $1.96\sigma$ exclusion boundary.}
			\label{fig:log_likelihood_ratio_kappa}
		\end{figure}

		\begin{table}[t]
			\centering
			\caption{Constraints on $\kappa_\lambda$ using different selection methods. We consider a luminosity of $\mathcal{L} = \text{300 fb}^{-1}$ for the 14-TeV LHC.}
			\label{tab:kappa_constraint_summary}
			\begin{tabular}{lcccccc}
			\hline\hline
										  &  \multicolumn{6}{c}{Expected Constraints}                                          \\
			POI							  & \multicolumn{3}{c}{~~Signal strength $\mu$~~}  &  \multicolumn{3}{c}{~~Self-coupling $\kappa_\lambda$~~}                 \\ \hline
			S/B selection                 & ~Lower~ & ~Upper~ & ~$\Delta \kappa_\lambda$~ &  ~Lower~ & ~Upper~ & ~$\Delta \kappa_\lambda$~ \\ \hline
			Cut-based with $\text{min-}\Delta R$ pairing & $-3.30$ & $10.38$ & $13.68$ &  $-3.27 $ & $10.33$ & $13.60$ \\
			Cut-based with \textsc{Spa-Net} pairing~	     & $-3.58$ & $10.47$ & $14.05$ &  $-3.55 $ & $10.43$ & $13.98$ \\
			DNN with $\text{min-}\Delta R$ pairing       & $-1.44$ & $8.49$  & $9.93 $ &  $-1.44 $ & $8.49$  & $9.93 $ \\
			DNN with \textsc{Spa-Net} pairing			 & $-1.72$ & $8.94$  & $10.66$ &  $-1.72 $ & $8.99$  & $10.71$ \\
			\textsc{Spa-Net} selection          		 & $-1.57$ & $7.49$  & $9.06 $ &  $-1.56 $ & $7.57$  & $9.13 $ \\
			\hline\hline
			\end{tabular}
		\end{table}

		When using the signal strength $\mu$ as the POI, Figure \ref{fig:CL_upper_limit} displays the upper limits of the $hh$ cross-section as a function of $\kappa_\lambda$.  The dip of the curves around $\kappa_\lambda = 2.45$ is due to the complete destructive interference~\cite{Barger:2014taa}.  On the other hand, when using $\kappa_\lambda$ as the POI, the profile likelihood ratio scan for $\kappa_\lambda$ is shown in Figure \ref{fig:log_likelihood_ratio_kappa}. All results are presented in Table~\ref{tab:kappa_constraint_summary}. The \textsc{Spa-Net} classifier shows a significant improvement in the upper bound of the $\kappa_\lambda$ constraints. Regardless of the selected POI, the \textsc{Spa-Net} classifier provides the strongest constraints on $\kappa_\lambda$. When using $\mu$ ($\kappa_\lambda$) as the POI, values of $\kappa_\lambda$ beyond $[-1.57, 7.49]$ ($[-1.56, 7.57]$) would be excluded at the 95\% CL ($1.96\sigma$) level.

		To achieve similar performance using the $\text{min-}\Delta R$ DNN classifier, the luminosity would need to be scaled to $\mathcal{L} \simeq \text{400 fb}^{-1}$. In this sense, the \textsc{Spa-Net} classifier offers a luminosity gain of $\mathcal{L} = \text{100 fb}^{-1}$. These results demonstrate that the \textsc{Spa-Net} classifier outperforms other methods in constraining Higgs self-coupling scale factor $\kappa_\lambda$.

		
\section{Conclusions}
\label{sec:conclusion}

    In this paper, we have utilized a novel deep neural network architecture called \textsc{Spa-Net} to improve the analysis for di-Higgs events in the $4b$ final state. By comparing \textsc{Spa-Net} with the traditional cut-based method, we have shown its better pairing efficiency in both resonant and non-resonant scenarios. Moreover, \textsc{Spa-Net} also outperforms the DNN classifier in the classification task.

    At the LHC with a 14-TeV C.M. energy and an integrated luminosity of 300 fb$^{-1}$, \textsc{Spa-Net} selection allows us to establish 95\% C.L. cross-section upper limits  in resonant analysis that are 10\% to 45\% stronger compared to DNN with $\text{min-}\Delta R$ pairing. In non-resonant analysis, the \textsc{Spa-Net} selection has provided 9\% stronger bounds on $\kappa_\lambda$ couplings when compared to the baseline method, the $\text{min-}\Delta R$ DNN selection. To achieve similar performance using the $\text{min-}\Delta R$ DNN selection, the luminosity would need to be scaled to $\mathcal{L} = \text{400 fb}^{-1}$. Therefore, the \textsc{Spa-Net} selection offers a luminosity gain of $\mathcal{L} = \text{100 fb}^{-1}$.

    The key difference between DNN and \textsc{Spa-Net} classifiers is in the input features. The \textsc{Spa-Net} classifier uses the embedding vectors as its input features. We have demonstrated that embedding vectors are related to high-level physical observables and they can capture the physical information from the events. This feature makes the embedding vectors well-suited for event classification tasks. In conclusion, our work highlights the remarkable potential of \textsc{Spa-Net} in di-Higgs event analysis. Its superior performance in both jet pairing and classification tasks holds significant promise for enhancing experimental searches in this domain.


\section*{Acknowledgments}
\label{sec:acknowledgments}

	We are grateful to Alexander Shmakov for the assistance with the \textsc{Spa-Net} package. C.-W. Chiang and F.-Y. Hsieh are supported in part by the National Science and Technology Council of Taiwan under Grant No. NSTC-111-2112-M-002-018-MY3. S.-C. Hsu is supported by the U.S. National Science Foundation grants No. number 2110963. Work at Argonne is supported in part by the U.S. Department of Energy under contract DE-AC02- 06CH11357. I. Low acknowledges the hospitality of the Phenomenology Group at National Taiwan University during the completion of this work.
	

\newpage

\appendix

\section{Hyperparameters}
\label{sec:hyperparameters}
	The neural network hyperparameters are optimized by the \verb|Optuna| hyperparameter optimization package. Table~\ref{tab:dnn_hyperparameter} and Table~\ref{tab:spanet_hyperparameter} provide the hyperparameters utilized in the DNN and \textsc{Spa-Net} training, respectively. 
	
	\begin{table}[htpb]
		\centering
		\caption{The hyperparameters used in DNN training. The results of different pairing methods are presented.}
		\label{tab:dnn_hyperparameter}
		\begin{tabular}{l|ll|ll}
		\hline \hline
		Parameter             & \multicolumn{2}{c|}{Resonant}            & \multicolumn{2}{c}{Non-resonant}         \\
							  & $\text{min-}\Delta R$ & \textsc{Spa-Net} & $\text{min-}\Delta R$ & \textsc{Spa-Net} \\ \hline
		Learning rate         & 0.000227              & 0.000262         & 0.00495               & 0.000948         \\
		Batch Size            & 512                   & 512              & 512                   & 512              \\
		Hidden Dimensionality & 256                   & 256              & 256                   & 256              \\
		Layer Count           & 5                     & 5                & 3                     & 2                \\
		\hline \hline
		\end{tabular}
	\end{table}

	\begin{table}[htpb]
		\centering
		\caption{The hyperparameters used in \textsc{Spa-Net} training.}
		\label{tab:spanet_hyperparameter}
		\begin{tabular}{l|l|l}
		\hline \hline
		Parameter                 & Resonant & Non-resonant \\ \hline
		Learning Rate             & 0.00049  & 0.00659      \\
		Training Epochs           & 50       & 50           \\
		Batch Size                & 2048     & 2048         \\
		Dropout                   & 0.061    & 0.0059       \\
		$L_2$ Gradient Clipping   & 0.445    & 0.425        \\
		$L_2$ Penalty             & 0.000382 & 0.000374     \\
		Hidden Dimensionality     & 256      & 32           \\
		Central Encoder Count     & 8        & 8            \\
		Branch Encoder Count      & 4        & 2            \\
		Classification Layers     & 1        & 1            \\ 
		Assignment Loss Scale     & 1        & 1            \\
		Classification Loss Scale & 1        & 1            \\
		\hline \hline
		\end{tabular}
	\end{table}

\bibliography{reference}

\begin{thebibliography}{49}%
\makeatletter
\providecommand \@ifxundefined [1]{%
 \@ifx{#1\undefined}
}%
\providecommand \@ifnum [1]{%
 \ifnum #1\expandafter \@firstoftwo
 \else \expandafter \@secondoftwo
 \fi
}%
\providecommand \@ifx [1]{%
 \ifx #1\expandafter \@firstoftwo
 \else \expandafter \@secondoftwo
 \fi
}%
\providecommand \natexlab [1]{#1}%
\providecommand \enquote  [1]{``#1''}%
\providecommand \bibnamefont  [1]{#1}%
\providecommand \bibfnamefont [1]{#1}%
\providecommand \citenamefont [1]{#1}%
\providecommand \href@noop [0]{\@secondoftwo}%
\providecommand \href [0]{\begingroup \@sanitize@url \@href}%
\providecommand \@href[1]{\@@startlink{#1}\@@href}%
\providecommand \@@href[1]{\endgroup#1\@@endlink}%
\providecommand \@sanitize@url [0]{\catcode `\\12\catcode `\$12\catcode
  `\&12\catcode `\#12\catcode `\^12\catcode `\_12\catcode `\%12\relax}%
\providecommand \@@startlink[1]{}%
\providecommand \@@endlink[0]{}%
\providecommand \url  [0]{\begingroup\@sanitize@url \@url }%
\providecommand \@url [1]{\endgroup\@href {#1}{\urlprefix }}%
\providecommand \urlprefix  [0]{URL }%
\providecommand \Eprint [0]{\href }%
\providecommand \doibase [0]{http://dx.doi.org/}%
\providecommand \selectlanguage [0]{\@gobble}%
\providecommand \bibinfo  [0]{\@secondoftwo}%
\providecommand \bibfield  [0]{\@secondoftwo}%
\providecommand \translation [1]{[#1]}%
\providecommand \BibitemOpen [0]{}%
\providecommand \bibitemStop [0]{}%
\providecommand \bibitemNoStop [0]{.\EOS\space}%
\providecommand \EOS [0]{\spacefactor3000\relax}%
\providecommand \BibitemShut  [1]{\csname bibitem#1\endcsname}%
\let\auto@bib@innerbib\@empty
\bibitem [{\citenamefont {Aad}\ \emph {et~al.}(2022{\natexlab{a}})\citenamefont
  {Aad} \emph {et~al.}}]{ATLAS:2022vkf}%
  \BibitemOpen
  \bibfield  {author} {\bibinfo {author} {\bibfnamefont {G.}~\bibnamefont
  {Aad}} \emph {et~al.} (\bibinfo {collaboration} {ATLAS}),\ }\href {\doibase
  10.1038/s41586-022-04893-w} {\bibfield  {journal} {\bibinfo  {journal}
  {Nature}\ }\textbf {\bibinfo {volume} {607}},\ \bibinfo {pages} {52}
  (\bibinfo {year} {2022}{\natexlab{a}})},\ \bibinfo {note} {[Erratum: Nature
  612, E24 (2022)]},\ \Eprint {http://arxiv.org/abs/2207.00092}
  {arXiv:2207.00092 [hep-ex]} \BibitemShut {NoStop}%
\bibitem [{\citenamefont {Tumasyan}\ \emph
  {et~al.}(2022{\natexlab{a}})\citenamefont {Tumasyan} \emph
  {et~al.}}]{CMS:2022dwd}%
  \BibitemOpen
  \bibfield  {author} {\bibinfo {author} {\bibfnamefont {A.}~\bibnamefont
  {Tumasyan}} \emph {et~al.} (\bibinfo {collaboration} {CMS}),\ }\href
  {\doibase 10.1038/s41586-022-04892-x} {\bibfield  {journal} {\bibinfo
  {journal} {Nature}\ }\textbf {\bibinfo {volume} {607}},\ \bibinfo {pages}
  {60} (\bibinfo {year} {2022}{\natexlab{a}})},\ \Eprint
  {http://arxiv.org/abs/2207.00043} {arXiv:2207.00043 [hep-ex]} \BibitemShut
  {NoStop}%
\bibitem [{\citenamefont {Sakharov}(1967)}]{Sakharov:1967dj}%
  \BibitemOpen
  \bibfield  {author} {\bibinfo {author} {\bibfnamefont {A.~D.}\ \bibnamefont
  {Sakharov}},\ }\href {\doibase 10.1070/PU1991v034n05ABEH002497} {\bibfield
  {journal} {\bibinfo  {journal} {Pisma Zh. Eksp. Teor. Fiz.}\ }\textbf
  {\bibinfo {volume} {5}},\ \bibinfo {pages} {32} (\bibinfo {year}
  {1967})}\BibitemShut {NoStop}%
\bibitem [{\citenamefont {Kuzmin}\ \emph {et~al.}(1985)\citenamefont {Kuzmin},
  \citenamefont {Rubakov},\ and\ \citenamefont {Shaposhnikov}}]{Kuzmin:1985mm}%
  \BibitemOpen
  \bibfield  {author} {\bibinfo {author} {\bibfnamefont {V.~A.}\ \bibnamefont
  {Kuzmin}}, \bibinfo {author} {\bibfnamefont {V.~A.}\ \bibnamefont {Rubakov}},
  \ and\ \bibinfo {author} {\bibfnamefont {M.~E.}\ \bibnamefont
  {Shaposhnikov}},\ }\href {\doibase 10.1016/0370-2693(85)91028-7} {\bibfield
  {journal} {\bibinfo  {journal} {Phys. Lett. B}\ }\textbf {\bibinfo {volume}
  {155}},\ \bibinfo {pages} {36} (\bibinfo {year} {1985})}\BibitemShut
  {NoStop}%
\bibitem [{\citenamefont {Cohen}\ \emph {et~al.}(1993)\citenamefont {Cohen},
  \citenamefont {Kaplan},\ and\ \citenamefont {Nelson}}]{Cohen:1993nk}%
  \BibitemOpen
  \bibfield  {author} {\bibinfo {author} {\bibfnamefont {A.~G.}\ \bibnamefont
  {Cohen}}, \bibinfo {author} {\bibfnamefont {D.~B.}\ \bibnamefont {Kaplan}}, \
  and\ \bibinfo {author} {\bibfnamefont {A.~E.}\ \bibnamefont {Nelson}},\
  }\href {\doibase 10.1146/annurev.ns.43.120193.000331} {\bibfield  {journal}
  {\bibinfo  {journal} {Ann. Rev. Nucl. Part. Sci.}\ }\textbf {\bibinfo
  {volume} {43}},\ \bibinfo {pages} {27} (\bibinfo {year} {1993})},\ \Eprint
  {http://arxiv.org/abs/hep-ph/9302210} {arXiv:hep-ph/9302210} \BibitemShut
  {NoStop}%
\bibitem [{\citenamefont {Kanemura}\ \emph {et~al.}(2005)\citenamefont
  {Kanemura}, \citenamefont {Okada},\ and\ \citenamefont
  {Senaha}}]{Kanemura:2004ch}%
  \BibitemOpen
  \bibfield  {author} {\bibinfo {author} {\bibfnamefont {S.}~\bibnamefont
  {Kanemura}}, \bibinfo {author} {\bibfnamefont {Y.}~\bibnamefont {Okada}}, \
  and\ \bibinfo {author} {\bibfnamefont {E.}~\bibnamefont {Senaha}},\ }\href
  {\doibase 10.1016/j.physletb.2004.12.004} {\bibfield  {journal} {\bibinfo
  {journal} {Phys. Lett. B}\ }\textbf {\bibinfo {volume} {606}},\ \bibinfo
  {pages} {361} (\bibinfo {year} {2005})},\ \Eprint
  {http://arxiv.org/abs/hep-ph/0411354} {arXiv:hep-ph/0411354} \BibitemShut
  {NoStop}%
\bibitem [{\citenamefont {Dawson}\ \emph {et~al.}(2015)\citenamefont {Dawson},
  \citenamefont {Ismail},\ and\ \citenamefont {Low}}]{Dawson:2015oha}%
  \BibitemOpen
  \bibfield  {author} {\bibinfo {author} {\bibfnamefont {S.}~\bibnamefont
  {Dawson}}, \bibinfo {author} {\bibfnamefont {A.}~\bibnamefont {Ismail}}, \
  and\ \bibinfo {author} {\bibfnamefont {I.}~\bibnamefont {Low}},\ }\href
  {\doibase 10.1103/PhysRevD.91.115008} {\bibfield  {journal} {\bibinfo
  {journal} {Phys. Rev. D}\ }\textbf {\bibinfo {volume} {91}},\ \bibinfo
  {pages} {115008} (\bibinfo {year} {2015})},\ \Eprint
  {http://arxiv.org/abs/1504.05596} {arXiv:1504.05596 [hep-ph]} \BibitemShut
  {NoStop}%
\bibitem [{\citenamefont {Chen}\ and\ \citenamefont
  {Low}(2014)}]{Chen:2014xra}%
  \BibitemOpen
  \bibfield  {author} {\bibinfo {author} {\bibfnamefont {C.-R.}\ \bibnamefont
  {Chen}}\ and\ \bibinfo {author} {\bibfnamefont {I.}~\bibnamefont {Low}},\
  }\href {\doibase 10.1103/PhysRevD.90.013018} {\bibfield  {journal} {\bibinfo
  {journal} {Phys. Rev. D}\ }\textbf {\bibinfo {volume} {90}},\ \bibinfo
  {pages} {013018} (\bibinfo {year} {2014})},\ \Eprint
  {http://arxiv.org/abs/1405.7040} {arXiv:1405.7040 [hep-ph]} \BibitemShut
  {NoStop}%
\bibitem [{\citenamefont {Fenton}\ \emph {et~al.}(2022)\citenamefont {Fenton},
  \citenamefont {Shmakov}, \citenamefont {Ho}, \citenamefont {Hsu},
  \citenamefont {Whiteson},\ and\ \citenamefont {Baldi}}]{PhysRevD.105.112008}%
  \BibitemOpen
  \bibfield  {author} {\bibinfo {author} {\bibfnamefont {M.~J.}\ \bibnamefont
  {Fenton}}, \bibinfo {author} {\bibfnamefont {A.}~\bibnamefont {Shmakov}},
  \bibinfo {author} {\bibfnamefont {T.-W.}\ \bibnamefont {Ho}}, \bibinfo
  {author} {\bibfnamefont {S.-C.}\ \bibnamefont {Hsu}}, \bibinfo {author}
  {\bibfnamefont {D.}~\bibnamefont {Whiteson}}, \ and\ \bibinfo {author}
  {\bibfnamefont {P.}~\bibnamefont {Baldi}},\ }\href {\doibase
  10.1103/PhysRevD.105.112008} {\bibfield  {journal} {\bibinfo  {journal}
  {Phys. Rev. D}\ }\textbf {\bibinfo {volume} {105}},\ \bibinfo {pages}
  {112008} (\bibinfo {year} {2022})}\BibitemShut {NoStop}%
\bibitem [{\citenamefont {Shmakov}\ \emph {et~al.}(2022)\citenamefont
  {Shmakov}, \citenamefont {Fenton}, \citenamefont {Ho}, \citenamefont {Hsu},
  \citenamefont {Whiteson},\ and\ \citenamefont
  {Baldi}}]{10.21468/SciPostPhys.12.5.178}%
  \BibitemOpen
  \bibfield  {author} {\bibinfo {author} {\bibfnamefont {A.}~\bibnamefont
  {Shmakov}}, \bibinfo {author} {\bibfnamefont {M.~J.}\ \bibnamefont {Fenton}},
  \bibinfo {author} {\bibfnamefont {T.-W.}\ \bibnamefont {Ho}}, \bibinfo
  {author} {\bibfnamefont {S.-C.}\ \bibnamefont {Hsu}}, \bibinfo {author}
  {\bibfnamefont {D.}~\bibnamefont {Whiteson}}, \ and\ \bibinfo {author}
  {\bibfnamefont {P.}~\bibnamefont {Baldi}},\ }\href {\doibase
  10.21468/SciPostPhys.12.5.178} {\bibfield  {journal} {\bibinfo  {journal}
  {SciPost Phys.}\ }\textbf {\bibinfo {volume} {12}},\ \bibinfo {pages} {178}
  (\bibinfo {year} {2022})}\BibitemShut {NoStop}%
\bibitem [{\citenamefont {Fenton}\ \emph {et~al.}(2023)\citenamefont {Fenton},
  \citenamefont {Shmakov}, \citenamefont {Okawa}, \citenamefont {Li},
  \citenamefont {Hsiao}, \citenamefont {Hsu}, \citenamefont {Whiteson},\ and\
  \citenamefont {Baldi}}]{Fenton:2023ikr}%
  \BibitemOpen
  \bibfield  {author} {\bibinfo {author} {\bibfnamefont {M.~J.}\ \bibnamefont
  {Fenton}}, \bibinfo {author} {\bibfnamefont {A.}~\bibnamefont {Shmakov}},
  \bibinfo {author} {\bibfnamefont {H.}~\bibnamefont {Okawa}}, \bibinfo
  {author} {\bibfnamefont {Y.}~\bibnamefont {Li}}, \bibinfo {author}
  {\bibfnamefont {K.-Y.}\ \bibnamefont {Hsiao}}, \bibinfo {author}
  {\bibfnamefont {S.-C.}\ \bibnamefont {Hsu}}, \bibinfo {author} {\bibfnamefont
  {D.}~\bibnamefont {Whiteson}}, \ and\ \bibinfo {author} {\bibfnamefont
  {P.}~\bibnamefont {Baldi}},\ }\href@noop {} {\  (\bibinfo {year} {2023})},\
  \Eprint {http://arxiv.org/abs/2309.01886} {arXiv:2309.01886 [hep-ex]}
  \BibitemShut {NoStop}%
\bibitem [{\citenamefont {Aaboud}\ \emph {et~al.}(2019)\citenamefont {Aaboud}
  \emph {et~al.}}]{ATLAS:2018rnh}%
  \BibitemOpen
  \bibfield  {author} {\bibinfo {author} {\bibfnamefont {M.}~\bibnamefont
  {Aaboud}} \emph {et~al.} (\bibinfo {collaboration} {ATLAS}),\ }\href
  {\doibase 10.1007/JHEP01(2019)030} {\bibfield  {journal} {\bibinfo  {journal}
  {JHEP}\ }\textbf {\bibinfo {volume} {01}},\ \bibinfo {pages} {030} (\bibinfo
  {year} {2019})},\ \Eprint {http://arxiv.org/abs/1804.06174} {arXiv:1804.06174
  [hep-ex]} \BibitemShut {NoStop}%
\bibitem [{\citenamefont {Aad}\ \emph {et~al.}(2022{\natexlab{b}})\citenamefont
  {Aad} \emph {et~al.}}]{ATLAS:2022hwc}%
  \BibitemOpen
  \bibfield  {author} {\bibinfo {author} {\bibfnamefont {G.}~\bibnamefont
  {Aad}} \emph {et~al.} (\bibinfo {collaboration} {ATLAS}),\ }\href {\doibase
  10.1103/PhysRevD.105.092002} {\bibfield  {journal} {\bibinfo  {journal}
  {Phys. Rev. D}\ }\textbf {\bibinfo {volume} {105}},\ \bibinfo {pages}
  {092002} (\bibinfo {year} {2022}{\natexlab{b}})},\ \Eprint
  {http://arxiv.org/abs/2202.07288} {arXiv:2202.07288 [hep-ex]} \BibitemShut
  {NoStop}%
\bibitem [{\citenamefont {Aad}\ \emph {et~al.}(2023{\natexlab{a}})\citenamefont
  {Aad} \emph {et~al.}}]{ATLAS:2023qzf}%
  \BibitemOpen
  \bibfield  {author} {\bibinfo {author} {\bibfnamefont {G.}~\bibnamefont
  {Aad}} \emph {et~al.} (\bibinfo {collaboration} {ATLAS}),\ }\href {\doibase
  10.1103/PhysRevD.108.052003} {\bibfield  {journal} {\bibinfo  {journal}
  {Phys. Rev. D}\ }\textbf {\bibinfo {volume} {108}},\ \bibinfo {pages}
  {052003} (\bibinfo {year} {2023}{\natexlab{a}})},\ \Eprint
  {http://arxiv.org/abs/2301.03212} {arXiv:2301.03212 [hep-ex]} \BibitemShut
  {NoStop}%
\bibitem [{\citenamefont {Sirunyan}\ \emph {et~al.}(2018)\citenamefont
  {Sirunyan} \emph {et~al.}}]{CMS:2018qmt}%
  \BibitemOpen
  \bibfield  {author} {\bibinfo {author} {\bibfnamefont {A.~M.}\ \bibnamefont
  {Sirunyan}} \emph {et~al.} (\bibinfo {collaboration} {CMS}),\ }\href
  {\doibase 10.1007/JHEP08(2018)152} {\bibfield  {journal} {\bibinfo  {journal}
  {JHEP}\ }\textbf {\bibinfo {volume} {08}},\ \bibinfo {pages} {152} (\bibinfo
  {year} {2018})},\ \Eprint {http://arxiv.org/abs/1806.03548} {arXiv:1806.03548
  [hep-ex]} \BibitemShut {NoStop}%
\bibitem [{\citenamefont {Tumasyan}\ \emph
  {et~al.}(2022{\natexlab{b}})\citenamefont {Tumasyan} \emph
  {et~al.}}]{CMS:2022cpr}%
  \BibitemOpen
  \bibfield  {author} {\bibinfo {author} {\bibfnamefont {A.}~\bibnamefont
  {Tumasyan}} \emph {et~al.} (\bibinfo {collaboration} {CMS}),\ }\href
  {\doibase 10.1103/PhysRevLett.129.081802} {\bibfield  {journal} {\bibinfo
  {journal} {Phys. Rev. Lett.}\ }\textbf {\bibinfo {volume} {129}},\ \bibinfo
  {pages} {081802} (\bibinfo {year} {2022}{\natexlab{b}})},\ \Eprint
  {http://arxiv.org/abs/2202.09617} {arXiv:2202.09617 [hep-ex]} \BibitemShut
  {NoStop}%
\bibitem [{\citenamefont {Amacker}\ \emph {et~al.}(2020)\citenamefont {Amacker}
  \emph {et~al.}}]{Amacker:2020bmn}%
  \BibitemOpen
  \bibfield  {author} {\bibinfo {author} {\bibfnamefont {J.}~\bibnamefont
  {Amacker}} \emph {et~al.},\ }\href {\doibase 10.1007/JHEP12(2020)115}
  {\bibfield  {journal} {\bibinfo  {journal} {JHEP}\ }\textbf {\bibinfo
  {volume} {12}},\ \bibinfo {pages} {115} (\bibinfo {year} {2020})},\ \Eprint
  {http://arxiv.org/abs/2004.04240} {arXiv:2004.04240 [hep-ph]} \BibitemShut
  {NoStop}%
\bibitem [{\citenamefont {Carena}\ \emph {et~al.}(2014)\citenamefont {Carena},
  \citenamefont {Low}, \citenamefont {Shah},\ and\ \citenamefont
  {Wagner}}]{Carena:2013ooa}%
  \BibitemOpen
  \bibfield  {author} {\bibinfo {author} {\bibfnamefont {M.}~\bibnamefont
  {Carena}}, \bibinfo {author} {\bibfnamefont {I.}~\bibnamefont {Low}},
  \bibinfo {author} {\bibfnamefont {N.~R.}\ \bibnamefont {Shah}}, \ and\
  \bibinfo {author} {\bibfnamefont {C.~E.~M.}\ \bibnamefont {Wagner}},\ }\href
  {\doibase 10.1007/JHEP04(2014)015} {\bibfield  {journal} {\bibinfo  {journal}
  {JHEP}\ }\textbf {\bibinfo {volume} {04}},\ \bibinfo {pages} {015} (\bibinfo
  {year} {2014})},\ \Eprint {http://arxiv.org/abs/1310.2248} {arXiv:1310.2248
  [hep-ph]} \BibitemShut {NoStop}%
\bibitem [{\citenamefont {Carena}\ \emph {et~al.}(2015)\citenamefont {Carena},
  \citenamefont {Haber}, \citenamefont {Low}, \citenamefont {Shah},\ and\
  \citenamefont {Wagner}}]{Carena:2014nza}%
  \BibitemOpen
  \bibfield  {author} {\bibinfo {author} {\bibfnamefont {M.}~\bibnamefont
  {Carena}}, \bibinfo {author} {\bibfnamefont {H.~E.}\ \bibnamefont {Haber}},
  \bibinfo {author} {\bibfnamefont {I.}~\bibnamefont {Low}}, \bibinfo {author}
  {\bibfnamefont {N.~R.}\ \bibnamefont {Shah}}, \ and\ \bibinfo {author}
  {\bibfnamefont {C.~E.~M.}\ \bibnamefont {Wagner}},\ }\href {\doibase
  10.1103/PhysRevD.91.035003} {\bibfield  {journal} {\bibinfo  {journal} {Phys.
  Rev. D}\ }\textbf {\bibinfo {volume} {91}},\ \bibinfo {pages} {035003}
  (\bibinfo {year} {2015})},\ \Eprint {http://arxiv.org/abs/1410.4969}
  {arXiv:1410.4969 [hep-ph]} \BibitemShut {NoStop}%
\bibitem [{\citenamefont {Carena}\ \emph {et~al.}(2016)\citenamefont {Carena},
  \citenamefont {Haber}, \citenamefont {Low}, \citenamefont {Shah},\ and\
  \citenamefont {Wagner}}]{Carena:2015moc}%
  \BibitemOpen
  \bibfield  {author} {\bibinfo {author} {\bibfnamefont {M.}~\bibnamefont
  {Carena}}, \bibinfo {author} {\bibfnamefont {H.~E.}\ \bibnamefont {Haber}},
  \bibinfo {author} {\bibfnamefont {I.}~\bibnamefont {Low}}, \bibinfo {author}
  {\bibfnamefont {N.~R.}\ \bibnamefont {Shah}}, \ and\ \bibinfo {author}
  {\bibfnamefont {C.~E.~M.}\ \bibnamefont {Wagner}},\ }\href {\doibase
  10.1103/PhysRevD.93.035013} {\bibfield  {journal} {\bibinfo  {journal} {Phys.
  Rev. D}\ }\textbf {\bibinfo {volume} {93}},\ \bibinfo {pages} {035013}
  (\bibinfo {year} {2016})},\ \Eprint {http://arxiv.org/abs/1510.09137}
  {arXiv:1510.09137 [hep-ph]} \BibitemShut {NoStop}%
\bibitem [{\citenamefont {Low}\ \emph {et~al.}(2022)\citenamefont {Low},
  \citenamefont {Shah},\ and\ \citenamefont {Wang}}]{Low:2020iua}%
  \BibitemOpen
  \bibfield  {author} {\bibinfo {author} {\bibfnamefont {I.}~\bibnamefont
  {Low}}, \bibinfo {author} {\bibfnamefont {N.~R.}\ \bibnamefont {Shah}}, \
  and\ \bibinfo {author} {\bibfnamefont {X.-P.}\ \bibnamefont {Wang}},\ }\href
  {\doibase 10.1103/PhysRevD.105.035009} {\bibfield  {journal} {\bibinfo
  {journal} {Phys. Rev. D}\ }\textbf {\bibinfo {volume} {105}},\ \bibinfo
  {pages} {035009} (\bibinfo {year} {2022})},\ \Eprint
  {http://arxiv.org/abs/2012.00773} {arXiv:2012.00773 [hep-ph]} \BibitemShut
  {NoStop}%
\bibitem [{\citenamefont {Chen}\ \emph {et~al.}(2022)\citenamefont {Chen},
  \citenamefont {Chiang},\ and\ \citenamefont {Low}}]{Chen:2022vac}%
  \BibitemOpen
  \bibfield  {author} {\bibinfo {author} {\bibfnamefont {T.-K.}\ \bibnamefont
  {Chen}}, \bibinfo {author} {\bibfnamefont {C.-W.}\ \bibnamefont {Chiang}}, \
  and\ \bibinfo {author} {\bibfnamefont {I.}~\bibnamefont {Low}},\ }\href
  {\doibase 10.1103/PhysRevD.105.075025} {\bibfield  {journal} {\bibinfo
  {journal} {Phys. Rev. D}\ }\textbf {\bibinfo {volume} {105}},\ \bibinfo
  {pages} {075025} (\bibinfo {year} {2022})},\ \Eprint
  {http://arxiv.org/abs/2202.02954} {arXiv:2202.02954 [hep-ph]} \BibitemShut
  {NoStop}%
\bibitem [{\citenamefont {Alwall}\ \emph {et~al.}(2014)\citenamefont {Alwall},
  \citenamefont {Frederix}, \citenamefont {Frixione}, \citenamefont {Hirschi},
  \citenamefont {Maltoni}, \citenamefont {Mattelaer}, \citenamefont {Shao},
  \citenamefont {Stelzer}, \citenamefont {Torrielli},\ and\ \citenamefont
  {Zaro}}]{Alwall:2014hca}%
  \BibitemOpen
  \bibfield  {author} {\bibinfo {author} {\bibfnamefont {J.}~\bibnamefont
  {Alwall}}, \bibinfo {author} {\bibfnamefont {R.}~\bibnamefont {Frederix}},
  \bibinfo {author} {\bibfnamefont {S.}~\bibnamefont {Frixione}}, \bibinfo
  {author} {\bibfnamefont {V.}~\bibnamefont {Hirschi}}, \bibinfo {author}
  {\bibfnamefont {F.}~\bibnamefont {Maltoni}}, \bibinfo {author} {\bibfnamefont
  {O.}~\bibnamefont {Mattelaer}}, \bibinfo {author} {\bibfnamefont {H.~S.}\
  \bibnamefont {Shao}}, \bibinfo {author} {\bibfnamefont {T.}~\bibnamefont
  {Stelzer}}, \bibinfo {author} {\bibfnamefont {P.}~\bibnamefont {Torrielli}},
  \ and\ \bibinfo {author} {\bibfnamefont {M.}~\bibnamefont {Zaro}},\ }\href
  {\doibase 10.1007/JHEP07(2014)079} {\bibfield  {journal} {\bibinfo  {journal}
  {JHEP}\ }\textbf {\bibinfo {volume} {07}},\ \bibinfo {pages} {079} (\bibinfo
  {year} {2014})},\ \Eprint {http://arxiv.org/abs/1405.0301} {arXiv:1405.0301
  [hep-ph]} \BibitemShut {NoStop}%
\bibitem [{\citenamefont {Ball}\ \emph {et~al.}(2013)\citenamefont {Ball} \emph
  {et~al.}}]{Ball:2012cx}%
  \BibitemOpen
  \bibfield  {author} {\bibinfo {author} {\bibfnamefont {R.~D.}\ \bibnamefont
  {Ball}} \emph {et~al.},\ }\href {\doibase 10.1016/j.nuclphysb.2012.10.003}
  {\bibfield  {journal} {\bibinfo  {journal} {Nucl. Phys. B}\ }\textbf
  {\bibinfo {volume} {867}},\ \bibinfo {pages} {244} (\bibinfo {year}
  {2013})},\ \Eprint {http://arxiv.org/abs/1207.1303} {arXiv:1207.1303
  [hep-ph]} \BibitemShut {NoStop}%
\bibitem [{\citenamefont {Sj\"ostrand}\ \emph {et~al.}(2015)\citenamefont
  {Sj\"ostrand}, \citenamefont {Ask}, \citenamefont {Christiansen},
  \citenamefont {Corke}, \citenamefont {Desai}, \citenamefont {Ilten},
  \citenamefont {Mrenna}, \citenamefont {Prestel}, \citenamefont {Rasmussen},\
  and\ \citenamefont {Skands}}]{Sjostrand:2014zea}%
  \BibitemOpen
  \bibfield  {author} {\bibinfo {author} {\bibfnamefont {T.}~\bibnamefont
  {Sj\"ostrand}}, \bibinfo {author} {\bibfnamefont {S.}~\bibnamefont {Ask}},
  \bibinfo {author} {\bibfnamefont {J.~R.}\ \bibnamefont {Christiansen}},
  \bibinfo {author} {\bibfnamefont {R.}~\bibnamefont {Corke}}, \bibinfo
  {author} {\bibfnamefont {N.}~\bibnamefont {Desai}}, \bibinfo {author}
  {\bibfnamefont {P.}~\bibnamefont {Ilten}}, \bibinfo {author} {\bibfnamefont
  {S.}~\bibnamefont {Mrenna}}, \bibinfo {author} {\bibfnamefont
  {S.}~\bibnamefont {Prestel}}, \bibinfo {author} {\bibfnamefont {C.~O.}\
  \bibnamefont {Rasmussen}}, \ and\ \bibinfo {author} {\bibfnamefont {P.~Z.}\
  \bibnamefont {Skands}},\ }\href {\doibase 10.1016/j.cpc.2015.01.024}
  {\bibfield  {journal} {\bibinfo  {journal} {Comput. Phys. Commun.}\ }\textbf
  {\bibinfo {volume} {191}},\ \bibinfo {pages} {159} (\bibinfo {year}
  {2015})},\ \Eprint {http://arxiv.org/abs/1410.3012} {arXiv:1410.3012
  [hep-ph]} \BibitemShut {NoStop}%
\bibitem [{\citenamefont {de~Favereau}\ \emph {et~al.}(2014)\citenamefont
  {de~Favereau}, \citenamefont {Delaere}, \citenamefont {Demin}, \citenamefont
  {Giammanco}, \citenamefont {Lema\^\i{}tre}, \citenamefont {Mertens},\ and\
  \citenamefont {Selvaggi}}]{deFavereau:2013fsa}%
  \BibitemOpen
  \bibfield  {author} {\bibinfo {author} {\bibfnamefont {J.}~\bibnamefont
  {de~Favereau}}, \bibinfo {author} {\bibfnamefont {C.}~\bibnamefont
  {Delaere}}, \bibinfo {author} {\bibfnamefont {P.}~\bibnamefont {Demin}},
  \bibinfo {author} {\bibfnamefont {A.}~\bibnamefont {Giammanco}}, \bibinfo
  {author} {\bibfnamefont {V.}~\bibnamefont {Lema\^\i{}tre}}, \bibinfo {author}
  {\bibfnamefont {A.}~\bibnamefont {Mertens}}, \ and\ \bibinfo {author}
  {\bibfnamefont {M.}~\bibnamefont {Selvaggi}} (\bibinfo {collaboration}
  {DELPHES 3}),\ }\href {\doibase 10.1007/JHEP02(2014)057} {\bibfield
  {journal} {\bibinfo  {journal} {JHEP}\ }\textbf {\bibinfo {volume} {02}},\
  \bibinfo {pages} {057} (\bibinfo {year} {2014})},\ \Eprint
  {http://arxiv.org/abs/1307.6346} {arXiv:1307.6346 [hep-ex]} \BibitemShut
  {NoStop}%
\bibitem [{\citenamefont {Cacciari}\ \emph {et~al.}(2012)\citenamefont
  {Cacciari}, \citenamefont {Salam},\ and\ \citenamefont
  {Soyez}}]{Cacciari:2011ma}%
  \BibitemOpen
  \bibfield  {author} {\bibinfo {author} {\bibfnamefont {M.}~\bibnamefont
  {Cacciari}}, \bibinfo {author} {\bibfnamefont {G.~P.}\ \bibnamefont {Salam}},
  \ and\ \bibinfo {author} {\bibfnamefont {G.}~\bibnamefont {Soyez}},\ }\href
  {\doibase 10.1140/epjc/s10052-012-1896-2} {\bibfield  {journal} {\bibinfo
  {journal} {Eur. Phys. J. C}\ }\textbf {\bibinfo {volume} {72}},\ \bibinfo
  {pages} {1896} (\bibinfo {year} {2012})},\ \Eprint
  {http://arxiv.org/abs/1111.6097} {arXiv:1111.6097 [hep-ph]} \BibitemShut
  {NoStop}%
\bibitem [{\citenamefont {Cacciari}\ \emph {et~al.}(2008)\citenamefont
  {Cacciari}, \citenamefont {Salam},\ and\ \citenamefont
  {Soyez}}]{Cacciari:2008gp}%
  \BibitemOpen
  \bibfield  {author} {\bibinfo {author} {\bibfnamefont {M.}~\bibnamefont
  {Cacciari}}, \bibinfo {author} {\bibfnamefont {G.~P.}\ \bibnamefont {Salam}},
  \ and\ \bibinfo {author} {\bibfnamefont {G.}~\bibnamefont {Soyez}},\ }\href
  {\doibase 10.1088/1126-6708/2008/04/063} {\bibfield  {journal} {\bibinfo
  {journal} {JHEP}\ }\textbf {\bibinfo {volume} {04}},\ \bibinfo {pages} {063}
  (\bibinfo {year} {2008})},\ \Eprint {http://arxiv.org/abs/0802.1189}
  {arXiv:0802.1189 [hep-ph]} \BibitemShut {NoStop}%
\bibitem [{\citenamefont {Eriksson}\ \emph {et~al.}(2010)\citenamefont
  {Eriksson}, \citenamefont {Rathsman},\ and\ \citenamefont
  {Stal}}]{Eriksson:2009ws}%
  \BibitemOpen
  \bibfield  {author} {\bibinfo {author} {\bibfnamefont {D.}~\bibnamefont
  {Eriksson}}, \bibinfo {author} {\bibfnamefont {J.}~\bibnamefont {Rathsman}},
  \ and\ \bibinfo {author} {\bibfnamefont {O.}~\bibnamefont {Stal}},\ }\href
  {\doibase 10.1016/j.cpc.2009.09.011} {\bibfield  {journal} {\bibinfo
  {journal} {Comput. Phys. Commun.}\ }\textbf {\bibinfo {volume} {181}},\
  \bibinfo {pages} {189} (\bibinfo {year} {2010})},\ \Eprint
  {http://arxiv.org/abs/0902.0851} {arXiv:0902.0851 [hep-ph]} \BibitemShut
  {NoStop}%
\bibitem [{\citenamefont {Bechtle}\ \emph {et~al.}(2010)\citenamefont
  {Bechtle}, \citenamefont {Brein}, \citenamefont {Heinemeyer}, \citenamefont
  {Weiglein},\ and\ \citenamefont {Williams}}]{Bechtle:2008jh}%
  \BibitemOpen
  \bibfield  {author} {\bibinfo {author} {\bibfnamefont {P.}~\bibnamefont
  {Bechtle}}, \bibinfo {author} {\bibfnamefont {O.}~\bibnamefont {Brein}},
  \bibinfo {author} {\bibfnamefont {S.}~\bibnamefont {Heinemeyer}}, \bibinfo
  {author} {\bibfnamefont {G.}~\bibnamefont {Weiglein}}, \ and\ \bibinfo
  {author} {\bibfnamefont {K.~E.}\ \bibnamefont {Williams}},\ }\href {\doibase
  10.1016/j.cpc.2009.09.003} {\bibfield  {journal} {\bibinfo  {journal}
  {Comput. Phys. Commun.}\ }\textbf {\bibinfo {volume} {181}},\ \bibinfo
  {pages} {138} (\bibinfo {year} {2010})},\ \Eprint
  {http://arxiv.org/abs/0811.4169} {arXiv:0811.4169 [hep-ph]} \BibitemShut
  {NoStop}%
\bibitem [{\citenamefont {Bechtle}\ \emph {et~al.}(2011)\citenamefont
  {Bechtle}, \citenamefont {Brein}, \citenamefont {Heinemeyer}, \citenamefont
  {Weiglein},\ and\ \citenamefont {Williams}}]{Bechtle:2011sb}%
  \BibitemOpen
  \bibfield  {author} {\bibinfo {author} {\bibfnamefont {P.}~\bibnamefont
  {Bechtle}}, \bibinfo {author} {\bibfnamefont {O.}~\bibnamefont {Brein}},
  \bibinfo {author} {\bibfnamefont {S.}~\bibnamefont {Heinemeyer}}, \bibinfo
  {author} {\bibfnamefont {G.}~\bibnamefont {Weiglein}}, \ and\ \bibinfo
  {author} {\bibfnamefont {K.~E.}\ \bibnamefont {Williams}},\ }\href {\doibase
  10.1016/j.cpc.2011.07.015} {\bibfield  {journal} {\bibinfo  {journal}
  {Comput. Phys. Commun.}\ }\textbf {\bibinfo {volume} {182}},\ \bibinfo
  {pages} {2605} (\bibinfo {year} {2011})},\ \Eprint
  {http://arxiv.org/abs/1102.1898} {arXiv:1102.1898 [hep-ph]} \BibitemShut
  {NoStop}%
\bibitem [{\citenamefont {Bechtle}\ \emph {et~al.}(2012)\citenamefont
  {Bechtle}, \citenamefont {Brein}, \citenamefont {Heinemeyer}, \citenamefont
  {Stal}, \citenamefont {Stefaniak}, \citenamefont {Weiglein},\ and\
  \citenamefont {Williams}}]{Bechtle:2012lvg}%
  \BibitemOpen
  \bibfield  {author} {\bibinfo {author} {\bibfnamefont {P.}~\bibnamefont
  {Bechtle}}, \bibinfo {author} {\bibfnamefont {O.}~\bibnamefont {Brein}},
  \bibinfo {author} {\bibfnamefont {S.}~\bibnamefont {Heinemeyer}}, \bibinfo
  {author} {\bibfnamefont {O.}~\bibnamefont {Stal}}, \bibinfo {author}
  {\bibfnamefont {T.}~\bibnamefont {Stefaniak}}, \bibinfo {author}
  {\bibfnamefont {G.}~\bibnamefont {Weiglein}}, \ and\ \bibinfo {author}
  {\bibfnamefont {K.}~\bibnamefont {Williams}},\ }\href {\doibase
  10.22323/1.156.0024} {\bibfield  {journal} {\bibinfo  {journal} {PoS}\
  }\textbf {\bibinfo {volume} {CHARGED2012}},\ \bibinfo {pages} {024} (\bibinfo
  {year} {2012})},\ \Eprint {http://arxiv.org/abs/1301.2345} {arXiv:1301.2345
  [hep-ph]} \BibitemShut {NoStop}%
\bibitem [{\citenamefont {Bechtle}\ \emph
  {et~al.}(2014{\natexlab{a}})\citenamefont {Bechtle}, \citenamefont {Brein},
  \citenamefont {Heinemeyer}, \citenamefont {St\r{a}l}, \citenamefont
  {Stefaniak}, \citenamefont {Weiglein},\ and\ \citenamefont
  {Williams}}]{Bechtle:2013wla}%
  \BibitemOpen
  \bibfield  {author} {\bibinfo {author} {\bibfnamefont {P.}~\bibnamefont
  {Bechtle}}, \bibinfo {author} {\bibfnamefont {O.}~\bibnamefont {Brein}},
  \bibinfo {author} {\bibfnamefont {S.}~\bibnamefont {Heinemeyer}}, \bibinfo
  {author} {\bibfnamefont {O.}~\bibnamefont {St\r{a}l}}, \bibinfo {author}
  {\bibfnamefont {T.}~\bibnamefont {Stefaniak}}, \bibinfo {author}
  {\bibfnamefont {G.}~\bibnamefont {Weiglein}}, \ and\ \bibinfo {author}
  {\bibfnamefont {K.~E.}\ \bibnamefont {Williams}},\ }\href {\doibase
  10.1140/epjc/s10052-013-2693-2} {\bibfield  {journal} {\bibinfo  {journal}
  {Eur. Phys. J. C}\ }\textbf {\bibinfo {volume} {74}},\ \bibinfo {pages}
  {2693} (\bibinfo {year} {2014}{\natexlab{a}})},\ \Eprint
  {http://arxiv.org/abs/1311.0055} {arXiv:1311.0055 [hep-ph]} \BibitemShut
  {NoStop}%
\bibitem [{\citenamefont {Bechtle}\ \emph {et~al.}(2015)\citenamefont
  {Bechtle}, \citenamefont {Heinemeyer}, \citenamefont {Stal}, \citenamefont
  {Stefaniak},\ and\ \citenamefont {Weiglein}}]{Bechtle:2015pma}%
  \BibitemOpen
  \bibfield  {author} {\bibinfo {author} {\bibfnamefont {P.}~\bibnamefont
  {Bechtle}}, \bibinfo {author} {\bibfnamefont {S.}~\bibnamefont {Heinemeyer}},
  \bibinfo {author} {\bibfnamefont {O.}~\bibnamefont {Stal}}, \bibinfo {author}
  {\bibfnamefont {T.}~\bibnamefont {Stefaniak}}, \ and\ \bibinfo {author}
  {\bibfnamefont {G.}~\bibnamefont {Weiglein}},\ }\href {\doibase
  10.1140/epjc/s10052-015-3650-z} {\bibfield  {journal} {\bibinfo  {journal}
  {Eur. Phys. J. C}\ }\textbf {\bibinfo {volume} {75}},\ \bibinfo {pages} {421}
  (\bibinfo {year} {2015})},\ \Eprint {http://arxiv.org/abs/1507.06706}
  {arXiv:1507.06706 [hep-ph]} \BibitemShut {NoStop}%
\bibitem [{\citenamefont {St\r{a}l}\ and\ \citenamefont
  {Stefaniak}(2013)}]{Stal:2013hwa}%
  \BibitemOpen
  \bibfield  {author} {\bibinfo {author} {\bibfnamefont {O.}~\bibnamefont
  {St\r{a}l}}\ and\ \bibinfo {author} {\bibfnamefont {T.}~\bibnamefont
  {Stefaniak}},\ }\href {\doibase 10.22323/1.180.0314} {\bibfield  {journal}
  {\bibinfo  {journal} {PoS}\ }\textbf {\bibinfo {volume} {EPS-HEP2013}},\
  \bibinfo {pages} {314} (\bibinfo {year} {2013})},\ \Eprint
  {http://arxiv.org/abs/1310.4039} {arXiv:1310.4039 [hep-ph]} \BibitemShut
  {NoStop}%
\bibitem [{\citenamefont {Bechtle}\ \emph
  {et~al.}(2014{\natexlab{b}})\citenamefont {Bechtle}, \citenamefont
  {Heinemeyer}, \citenamefont {St\r{a}l}, \citenamefont {Stefaniak},\ and\
  \citenamefont {Weiglein}}]{Bechtle:2013xfa}%
  \BibitemOpen
  \bibfield  {author} {\bibinfo {author} {\bibfnamefont {P.}~\bibnamefont
  {Bechtle}}, \bibinfo {author} {\bibfnamefont {S.}~\bibnamefont {Heinemeyer}},
  \bibinfo {author} {\bibfnamefont {O.}~\bibnamefont {St\r{a}l}}, \bibinfo
  {author} {\bibfnamefont {T.}~\bibnamefont {Stefaniak}}, \ and\ \bibinfo
  {author} {\bibfnamefont {G.}~\bibnamefont {Weiglein}},\ }\href {\doibase
  10.1140/epjc/s10052-013-2711-4} {\bibfield  {journal} {\bibinfo  {journal}
  {Eur. Phys. J. C}\ }\textbf {\bibinfo {volume} {74}},\ \bibinfo {pages}
  {2711} (\bibinfo {year} {2014}{\natexlab{b}})},\ \Eprint
  {http://arxiv.org/abs/1305.1933} {arXiv:1305.1933 [hep-ph]} \BibitemShut
  {NoStop}%
\bibitem [{\citenamefont {Bechtle}\ \emph
  {et~al.}(2014{\natexlab{c}})\citenamefont {Bechtle}, \citenamefont
  {Heinemeyer}, \citenamefont {St\r{a}l}, \citenamefont {Stefaniak},\ and\
  \citenamefont {Weiglein}}]{Bechtle:2014ewa}%
  \BibitemOpen
  \bibfield  {author} {\bibinfo {author} {\bibfnamefont {P.}~\bibnamefont
  {Bechtle}}, \bibinfo {author} {\bibfnamefont {S.}~\bibnamefont {Heinemeyer}},
  \bibinfo {author} {\bibfnamefont {O.}~\bibnamefont {St\r{a}l}}, \bibinfo
  {author} {\bibfnamefont {T.}~\bibnamefont {Stefaniak}}, \ and\ \bibinfo
  {author} {\bibfnamefont {G.}~\bibnamefont {Weiglein}},\ }\href {\doibase
  10.1007/JHEP11(2014)039} {\bibfield  {journal} {\bibinfo  {journal} {JHEP}\
  }\textbf {\bibinfo {volume} {11}},\ \bibinfo {pages} {039} (\bibinfo {year}
  {2014}{\natexlab{c}})},\ \Eprint {http://arxiv.org/abs/1403.1582}
  {arXiv:1403.1582 [hep-ph]} \BibitemShut {NoStop}%
\bibitem [{\citenamefont {Bechtle}\ \emph {et~al.}(2021)\citenamefont
  {Bechtle}, \citenamefont {Heinemeyer}, \citenamefont {Klingl}, \citenamefont
  {Stefaniak}, \citenamefont {Weiglein},\ and\ \citenamefont
  {Wittbrodt}}]{Bechtle:2020uwn}%
  \BibitemOpen
  \bibfield  {author} {\bibinfo {author} {\bibfnamefont {P.}~\bibnamefont
  {Bechtle}}, \bibinfo {author} {\bibfnamefont {S.}~\bibnamefont {Heinemeyer}},
  \bibinfo {author} {\bibfnamefont {T.}~\bibnamefont {Klingl}}, \bibinfo
  {author} {\bibfnamefont {T.}~\bibnamefont {Stefaniak}}, \bibinfo {author}
  {\bibfnamefont {G.}~\bibnamefont {Weiglein}}, \ and\ \bibinfo {author}
  {\bibfnamefont {J.}~\bibnamefont {Wittbrodt}},\ }\href {\doibase
  10.1140/epjc/s10052-021-08942-y} {\bibfield  {journal} {\bibinfo  {journal}
  {Eur. Phys. J. C}\ }\textbf {\bibinfo {volume} {81}},\ \bibinfo {pages} {145}
  (\bibinfo {year} {2021})},\ \Eprint {http://arxiv.org/abs/2012.09197}
  {arXiv:2012.09197 [hep-ph]} \BibitemShut {NoStop}%
\bibitem [{\citenamefont {Artoisenet}\ \emph {et~al.}(2013)\citenamefont
  {Artoisenet}, \citenamefont {Frederix}, \citenamefont {Mattelaer},\ and\
  \citenamefont {Rietkerk}}]{Artoisenet:2012st}%
  \BibitemOpen
  \bibfield  {author} {\bibinfo {author} {\bibfnamefont {P.}~\bibnamefont
  {Artoisenet}}, \bibinfo {author} {\bibfnamefont {R.}~\bibnamefont
  {Frederix}}, \bibinfo {author} {\bibfnamefont {O.}~\bibnamefont {Mattelaer}},
  \ and\ \bibinfo {author} {\bibfnamefont {R.}~\bibnamefont {Rietkerk}},\
  }\href {\doibase 10.1007/JHEP03(2013)015} {\bibfield  {journal} {\bibinfo
  {journal} {JHEP}\ }\textbf {\bibinfo {volume} {03}},\ \bibinfo {pages} {015}
  (\bibinfo {year} {2013})},\ \Eprint {http://arxiv.org/abs/1212.3460}
  {arXiv:1212.3460 [hep-ph]} \BibitemShut {NoStop}%
\bibitem [{ATL(2016)}]{ATL-PHYS-PUB-2016-012}%
  \BibitemOpen
  \href {https://cds.cern.ch/record/2160731} {\emph {\bibinfo {title}
  {{Optimisation of the ATLAS $b$-tagging performance for the 2016 LHC
  Run}}}},\ \bibinfo {type} {Tech. Rep.}\ (\bibinfo  {institution} {CERN},\
  \bibinfo {address} {Geneva},\ \bibinfo {year} {2016})\ \bibinfo {note} {all
  figures including auxiliary figures are available at
  https://atlas.web.cern.ch/Atlas/GROUPS/PHYSICS/PUBNOTES/ATL-PHYS-PUB-2016-012}\BibitemShut
  {NoStop}%
\bibitem [{\citenamefont {Aad}\ \emph {et~al.}(2016)\citenamefont {Aad} \emph
  {et~al.}}]{ATLAS:2015thz}%
  \BibitemOpen
  \bibfield  {author} {\bibinfo {author} {\bibfnamefont {G.}~\bibnamefont
  {Aad}} \emph {et~al.} (\bibinfo {collaboration} {ATLAS}),\ }\href {\doibase
  10.1088/1748-0221/11/04/P04008} {\bibfield  {journal} {\bibinfo  {journal}
  {JINST}\ }\textbf {\bibinfo {volume} {11}},\ \bibinfo {pages} {P04008}
  (\bibinfo {year} {2016})},\ \Eprint {http://arxiv.org/abs/1512.01094}
  {arXiv:1512.01094 [hep-ex]} \BibitemShut {NoStop}%
\bibitem [{\citenamefont {Aad}\ \emph {et~al.}(2023{\natexlab{b}})\citenamefont
  {Aad} \emph {et~al.}}]{ATLAS:2022qxm}%
  \BibitemOpen
  \bibfield  {author} {\bibinfo {author} {\bibfnamefont {G.}~\bibnamefont
  {Aad}} \emph {et~al.} (\bibinfo {collaboration} {ATLAS}),\ }\href {\doibase
  10.1140/epjc/s10052-023-11699-1} {\bibfield  {journal} {\bibinfo  {journal}
  {Eur. Phys. J. C}\ }\textbf {\bibinfo {volume} {83}},\ \bibinfo {pages} {681}
  (\bibinfo {year} {2023}{\natexlab{b}})},\ \Eprint
  {http://arxiv.org/abs/2211.16345} {arXiv:2211.16345 [physics.data-an]}
  \BibitemShut {NoStop}%
\bibitem [{\citenamefont {Abadi}\ \emph {et~al.}(2015)\citenamefont {Abadi},
  \citenamefont {Agarwal}, \citenamefont {Barham}, \citenamefont {Brevdo},
  \citenamefont {Chen}, \citenamefont {Citro}, \citenamefont {Corrado},
  \citenamefont {Davis}, \citenamefont {Dean}, \citenamefont {Devin},
  \citenamefont {Ghemawat}, \citenamefont {Goodfellow}, \citenamefont {Harp},
  \citenamefont {Irving}, \citenamefont {Isard}, \citenamefont {Jia},
  \citenamefont {Jozefowicz}, \citenamefont {Kaiser}, \citenamefont {Kudlur},
  \citenamefont {Levenberg}, \citenamefont {Man\'{e}}, \citenamefont {Monga},
  \citenamefont {Moore}, \citenamefont {Murray}, \citenamefont {Olah},
  \citenamefont {Schuster}, \citenamefont {Shlens}, \citenamefont {Steiner},
  \citenamefont {Sutskever}, \citenamefont {Talwar}, \citenamefont {Tucker},
  \citenamefont {Vanhoucke}, \citenamefont {Vasudevan}, \citenamefont
  {Vi\'{e}gas}, \citenamefont {Vinyals}, \citenamefont {Warden}, \citenamefont
  {Wattenberg}, \citenamefont {Wicke}, \citenamefont {Yu},\ and\ \citenamefont
  {Zheng}}]{tensorflow2015-whitepaper}%
  \BibitemOpen
  \bibfield  {author} {\bibinfo {author} {\bibfnamefont {M.}~\bibnamefont
  {Abadi}}, \bibinfo {author} {\bibfnamefont {A.}~\bibnamefont {Agarwal}},
  \bibinfo {author} {\bibfnamefont {P.}~\bibnamefont {Barham}}, \bibinfo
  {author} {\bibfnamefont {E.}~\bibnamefont {Brevdo}}, \bibinfo {author}
  {\bibfnamefont {Z.}~\bibnamefont {Chen}}, \bibinfo {author} {\bibfnamefont
  {C.}~\bibnamefont {Citro}}, \bibinfo {author} {\bibfnamefont {G.~S.}\
  \bibnamefont {Corrado}}, \bibinfo {author} {\bibfnamefont {A.}~\bibnamefont
  {Davis}}, \bibinfo {author} {\bibfnamefont {J.}~\bibnamefont {Dean}},
  \bibinfo {author} {\bibfnamefont {M.}~\bibnamefont {Devin}}, \bibinfo
  {author} {\bibfnamefont {S.}~\bibnamefont {Ghemawat}}, \bibinfo {author}
  {\bibfnamefont {I.}~\bibnamefont {Goodfellow}}, \bibinfo {author}
  {\bibfnamefont {A.}~\bibnamefont {Harp}}, \bibinfo {author} {\bibfnamefont
  {G.}~\bibnamefont {Irving}}, \bibinfo {author} {\bibfnamefont
  {M.}~\bibnamefont {Isard}}, \bibinfo {author} {\bibfnamefont
  {Y.}~\bibnamefont {Jia}}, \bibinfo {author} {\bibfnamefont {R.}~\bibnamefont
  {Jozefowicz}}, \bibinfo {author} {\bibfnamefont {L.}~\bibnamefont {Kaiser}},
  \bibinfo {author} {\bibfnamefont {M.}~\bibnamefont {Kudlur}}, \bibinfo
  {author} {\bibfnamefont {J.}~\bibnamefont {Levenberg}}, \bibinfo {author}
  {\bibfnamefont {D.}~\bibnamefont {Man\'{e}}}, \bibinfo {author}
  {\bibfnamefont {R.}~\bibnamefont {Monga}}, \bibinfo {author} {\bibfnamefont
  {S.}~\bibnamefont {Moore}}, \bibinfo {author} {\bibfnamefont
  {D.}~\bibnamefont {Murray}}, \bibinfo {author} {\bibfnamefont
  {C.}~\bibnamefont {Olah}}, \bibinfo {author} {\bibfnamefont {M.}~\bibnamefont
  {Schuster}}, \bibinfo {author} {\bibfnamefont {J.}~\bibnamefont {Shlens}},
  \bibinfo {author} {\bibfnamefont {B.}~\bibnamefont {Steiner}}, \bibinfo
  {author} {\bibfnamefont {I.}~\bibnamefont {Sutskever}}, \bibinfo {author}
  {\bibfnamefont {K.}~\bibnamefont {Talwar}}, \bibinfo {author} {\bibfnamefont
  {P.}~\bibnamefont {Tucker}}, \bibinfo {author} {\bibfnamefont
  {V.}~\bibnamefont {Vanhoucke}}, \bibinfo {author} {\bibfnamefont
  {V.}~\bibnamefont {Vasudevan}}, \bibinfo {author} {\bibfnamefont
  {F.}~\bibnamefont {Vi\'{e}gas}}, \bibinfo {author} {\bibfnamefont
  {O.}~\bibnamefont {Vinyals}}, \bibinfo {author} {\bibfnamefont
  {P.}~\bibnamefont {Warden}}, \bibinfo {author} {\bibfnamefont
  {M.}~\bibnamefont {Wattenberg}}, \bibinfo {author} {\bibfnamefont
  {M.}~\bibnamefont {Wicke}}, \bibinfo {author} {\bibfnamefont
  {Y.}~\bibnamefont {Yu}}, \ and\ \bibinfo {author} {\bibfnamefont
  {X.}~\bibnamefont {Zheng}},\ }\href {https://www.tensorflow.org/} {\enquote
  {\bibinfo {title} {{TensorFlow}: Large-scale machine learning on
  heterogeneous systems},}\ } (\bibinfo {year} {2015}),\ \bibinfo {note}
  {software available from tensorflow.org}\BibitemShut {NoStop}%
\bibitem [{\citenamefont {Akiba}\ \emph {et~al.}(2019)\citenamefont {Akiba},
  \citenamefont {Sano}, \citenamefont {Yanase}, \citenamefont {Ohta},\ and\
  \citenamefont {Koyama}}]{akiba2019optuna}%
  \BibitemOpen
  \bibfield  {author} {\bibinfo {author} {\bibfnamefont {T.}~\bibnamefont
  {Akiba}}, \bibinfo {author} {\bibfnamefont {S.}~\bibnamefont {Sano}},
  \bibinfo {author} {\bibfnamefont {T.}~\bibnamefont {Yanase}}, \bibinfo
  {author} {\bibfnamefont {T.}~\bibnamefont {Ohta}}, \ and\ \bibinfo {author}
  {\bibfnamefont {M.}~\bibnamefont {Koyama}},\ }\href@noop {} {\enquote
  {\bibinfo {title} {Optuna: A next-generation hyperparameter optimization
  framework},}\ } (\bibinfo {year} {2019}),\ \Eprint
  {http://arxiv.org/abs/1907.10902} {arXiv:1907.10902 [cs.LG]} \BibitemShut
  {NoStop}%
\bibitem [{\citenamefont {Pedregosa}\ \emph {et~al.}(2011)\citenamefont
  {Pedregosa}, \citenamefont {Varoquaux}, \citenamefont {Gramfort},
  \citenamefont {Michel}, \citenamefont {Thirion}, \citenamefont {Grisel},
  \citenamefont {Blondel}, \citenamefont {Prettenhofer}, \citenamefont {Weiss},
  \citenamefont {Dubourg}, \citenamefont {Vanderplas}, \citenamefont {Passos},
  \citenamefont {Cournapeau}, \citenamefont {Brucher}, \citenamefont {Perrot},\
  and\ \citenamefont {Duchesnay}}]{scikit-learn}%
  \BibitemOpen
  \bibfield  {author} {\bibinfo {author} {\bibfnamefont {F.}~\bibnamefont
  {Pedregosa}}, \bibinfo {author} {\bibfnamefont {G.}~\bibnamefont
  {Varoquaux}}, \bibinfo {author} {\bibfnamefont {A.}~\bibnamefont {Gramfort}},
  \bibinfo {author} {\bibfnamefont {V.}~\bibnamefont {Michel}}, \bibinfo
  {author} {\bibfnamefont {B.}~\bibnamefont {Thirion}}, \bibinfo {author}
  {\bibfnamefont {O.}~\bibnamefont {Grisel}}, \bibinfo {author} {\bibfnamefont
  {M.}~\bibnamefont {Blondel}}, \bibinfo {author} {\bibfnamefont
  {P.}~\bibnamefont {Prettenhofer}}, \bibinfo {author} {\bibfnamefont
  {R.}~\bibnamefont {Weiss}}, \bibinfo {author} {\bibfnamefont
  {V.}~\bibnamefont {Dubourg}}, \bibinfo {author} {\bibfnamefont
  {J.}~\bibnamefont {Vanderplas}}, \bibinfo {author} {\bibfnamefont
  {A.}~\bibnamefont {Passos}}, \bibinfo {author} {\bibfnamefont
  {D.}~\bibnamefont {Cournapeau}}, \bibinfo {author} {\bibfnamefont
  {M.}~\bibnamefont {Brucher}}, \bibinfo {author} {\bibfnamefont
  {M.}~\bibnamefont {Perrot}}, \ and\ \bibinfo {author} {\bibfnamefont
  {E.}~\bibnamefont {Duchesnay}},\ }\href@noop {} {\bibfield  {journal}
  {\bibinfo  {journal} {Journal of Machine Learning Research}\ }\textbf
  {\bibinfo {volume} {12}},\ \bibinfo {pages} {2825} (\bibinfo {year}
  {2011})}\BibitemShut {NoStop}%
\bibitem [{\citenamefont {Heinrich}\ \emph {et~al.}()\citenamefont {Heinrich},
  \citenamefont {Feickert},\ and\ \citenamefont {Stark}}]{pyhf}%
  \BibitemOpen
  \bibfield  {author} {\bibinfo {author} {\bibfnamefont {L.}~\bibnamefont
  {Heinrich}}, \bibinfo {author} {\bibfnamefont {M.}~\bibnamefont {Feickert}},
  \ and\ \bibinfo {author} {\bibfnamefont {G.}~\bibnamefont {Stark}},\ }\href
  {\doibase 10.5281/zenodo.1169739} {\enquote {\bibinfo {title} {{pyhf:
  v0.7.3}},}\ }\bibinfo {note}
  {Https://github.com/scikit-hep/pyhf/releases/tag/v0.7.3}\BibitemShut
  {NoStop}%
\bibitem [{\citenamefont {Heinrich}\ \emph {et~al.}(2021)\citenamefont
  {Heinrich}, \citenamefont {Feickert}, \citenamefont {Stark},\ and\
  \citenamefont {Cranmer}}]{pyhf_joss}%
  \BibitemOpen
  \bibfield  {author} {\bibinfo {author} {\bibfnamefont {L.}~\bibnamefont
  {Heinrich}}, \bibinfo {author} {\bibfnamefont {M.}~\bibnamefont {Feickert}},
  \bibinfo {author} {\bibfnamefont {G.}~\bibnamefont {Stark}}, \ and\ \bibinfo
  {author} {\bibfnamefont {K.}~\bibnamefont {Cranmer}},\ }\href {\doibase
  10.21105/joss.02823} {\bibfield  {journal} {\bibinfo  {journal} {Journal of
  Open Source Software}\ }\textbf {\bibinfo {volume} {6}},\ \bibinfo {pages}
  {2823} (\bibinfo {year} {2021})}\BibitemShut {NoStop}%
\bibitem [{\citenamefont {Read}(2002)}]{Read:2002hq}%
  \BibitemOpen
  \bibfield  {author} {\bibinfo {author} {\bibfnamefont {A.~L.}\ \bibnamefont
  {Read}},\ }\href {\doibase 10.1088/0954-3899/28/10/313} {\bibfield  {journal}
  {\bibinfo  {journal} {J. Phys. G}\ }\textbf {\bibinfo {volume} {28}},\
  \bibinfo {pages} {2693} (\bibinfo {year} {2002})}\BibitemShut {NoStop}%
\bibitem [{\citenamefont {Barger}\ \emph {et~al.}(2015)\citenamefont {Barger},
  \citenamefont {Everett}, \citenamefont {Jackson}, \citenamefont {Peterson},\
  and\ \citenamefont {Shaughnessy}}]{Barger:2014taa}%
  \BibitemOpen
  \bibfield  {author} {\bibinfo {author} {\bibfnamefont {V.}~\bibnamefont
  {Barger}}, \bibinfo {author} {\bibfnamefont {L.~L.}\ \bibnamefont {Everett}},
  \bibinfo {author} {\bibfnamefont {C.~B.}\ \bibnamefont {Jackson}}, \bibinfo
  {author} {\bibfnamefont {A.~D.}\ \bibnamefont {Peterson}}, \ and\ \bibinfo
  {author} {\bibfnamefont {G.}~\bibnamefont {Shaughnessy}},\ }\href {\doibase
  10.1103/PhysRevLett.114.011801} {\bibfield  {journal} {\bibinfo  {journal}
  {Phys. Rev. Lett.}\ }\textbf {\bibinfo {volume} {114}},\ \bibinfo {pages}
  {011801} (\bibinfo {year} {2015})},\ \Eprint {http://arxiv.org/abs/1408.0003}
  {arXiv:1408.0003 [hep-ph]} \BibitemShut {NoStop}%
\end{thebibliography}%

\end{document}